\documentclass[11pt,a4paper]{article}
\pdfoutput=1
\usepackage{jheppub}

\usepackage{epsfig}
\usepackage{latexsym}
\usepackage{amsfonts}
\usepackage{amsmath}
\usepackage{amsthm}
\usepackage{amssymb}
\usepackage{amsbsy}
\usepackage{multirow}

\newcommand{\CC}{\mathbb{C}} 
\newcommand{\RR}{\mathbb{R}} 
\newcommand{\ZZ}{\mathbb{Z}} 
\newcommand{\NN}{\mathbb{N}} 

\def\tr         {{\rm  tr}}
\def\cala         {{\cal A}}

\def\calm         {{\cal M}}

\def\calo         {{\cal O}}

\newsavebox{\uuunit}
\sbox{\uuunit}
    {\setlength{\unitlength}{0.825em}
     \begin{picture}(0.6,0.7)
        \thinlines
        \put(0,0){\line(1,0){0.5}}
        \put(0.15,0){\line(0,1){0.7}}
        \put(0.35,0){\line(0,1){0.8}}
       \multiput(0.3,0.8)(-0.04,-0.02){12}{\rule{0.5pt}{0.5pt}}
     \end {picture}}

\def\be{\begin{equation}}
\def\ee{\end{equation}}
\def\bea{\begin{eqnarray}}
\def\eea{\end{eqnarray}}

\def\h{\eta}

\def\d{\delta}
\def\e{\epsilon}

\def\l{\lambda}
\def\L{\Lambda}

\def\f{\phi}

\def\m{\mu}
\def\n{\nu}
\def\o{\omega}

\def\p{\pi}
\def\r{\rho}

\def\s{\sigma}

\def\t{\tau}

\def\sF{{{ F}\!\!\!\!\hskip.8pt\hbox{\raise1pt\hbox{/}}\,}}
\def\som{{{ \omega}\!\!\!\!\hskip.8pt\hbox{\raise1pt\hbox{/}}\,}}
\def\sJ{{{\rm J}\!\!\!\!\hskip.8pt\hbox{\raise1pt\hbox{/}}\,}}

\def\F{\Phi}
\def\pa{\partial}

\def\to{\rightarrow}
\def\nonu{\nonumber \\{}}
\def\half{{1 \over 2}}

\theoremstyle{definition}
\newtheorem*{remark}{Remark}
\newcommand{\rem}{\begin{remark}}
\newcommand{\erem}{\end{remark}}

\title{Multi-centered AdS$_3$ solutions from Virasoro conformal blocks}
\author[a,b]{Ond\v{r}ej Hul\'{i}k}
\author[c]{Tom\'{a}\v{s} Proch\'{a}zka}
\author[a]{Joris Raeymaekers}
\affiliation[a]{Institute of Physics of the ASCR,  Na Slovance 2, 182 21 Prague 8, Czech Republic.}
\affiliation[b]{Institute of Particle Physics and Nuclear Physics, Faculty of Mathematics and Physics, \\Charles University, V Hole\v{s}ovi\v{c}k\'{a}ch 2, 180 00 Prague 8, Czech Republic.}
\affiliation[c]{Arnold Sommerfeld Center for Theoretical Physics, Ludwig Maximilian University of Munich, Theresienstr. 37, D-80333 M\"unchen, Germany}
\emailAdd{ondra.hulik@gmail.com}\emailAdd{Tomas.Prochazka@lmu.de}\emailAdd{joris@fzu.cz}

\abstract{We revisit the construction of multi-centered solutions in three-dimensional anti-de Sitter gravity
in the light of the recently discovered connection between particle worldlines and classical Virasoro conformal blocks.
We focus on multi-centered solutions which represent the backreaction of point masses moving on helical geodesics in global AdS$_3$,
and argue that their construction
reduces to a problem in Liouville theory on the disk with Zamolodchikov-Zamolodchikov boundary condition. In order to construct the solution one needs to solve  a certain monodromy problem which we argue is solved by a vacuum classical conformal block on the sphere in a particular channel. In this way we construct multi-centered gravity solutions by using conformal blocks special functions. We show that our solutions represent  left-right asymmetric configurations of operator insertions in the dual CFT. We also provide a check of our arguments in an example and comment on other types of solutions.}
\arxivnumber{1612.03879}
\begin{document}
\maketitle
\section{Introduction and summary}
The AdS/CFT correspondence \cite{Maldacena:1997re} has proved to be a powerful tool to explore aspects of quantum gravity in anti-de-Sitter backgrounds. Recent investigations have focused on
the important question of how local physics in the bulk is encoded in properties of the dual CFT.  Natural localized objects to study holographically are line defects arising from adding point particles to the bulk, which have recently been shown
to be intimately linked to conformal blocks in the dual CFT \cite{Hartman:2013mia},\cite{Faulkner:2013yia},\cite{Fitzpatrick:2014vua},\cite{Hijano:2015rla} (related work appears in \cite{Hijano:2015zsa}-\cite{Alkalaev:2016rjl}). In the case of AdS$_3$, which we will focus on in this paper, it was shown that Virasoro conformal blocks at large central charge can be computed from the  action of configurations of point particles in the bulk AdS$_3$.

So far, this relation has been explored mostly in a so-called `heavy-light' approximation which on the bulk side means that only one of the bulk particles is allowed to backreact on the geometry while the other particles are treated as light probes. To go beyond the heavy-light approximation one has to consider fully backreacted  multi-particle  solutions, which will be the focus of this work. Although such multi-centered solutions in AdS$_3$ have been discussed in the literature  starting with 
\cite{Deser:1983tn},\cite{Deser:1983nh}, their relation  to   CFT conformal blocks has so far  not been elucidated. In this work we will describe  a precise connection between  multi-particle solutions in Lorentzian AdS$_3$ and conformal blocks in  Euclidean 2D CFT.
We will show that constructing a class of multi-centered gravity solutions amounts to  solving a particular monodromy problem, which also arises in the computation of a  conformal block for a specific class of correlators at large central charge,
using the monodromy method \cite{Zam0} (see also \cite{Harlow:2011ny}, \cite{Hartman:2013mia}).
Therefore, the fact that  the conformal block exists, together with its known properties at large central charge,
guarantees the existence of a solution to the gravity problem, even though we don't know the explicit  solution  in most cases.
The  multi-centered gravity solution can be constructed, in principle, from the conformal block, using it as a kind of special function.

\begin{figure}\begin{center}\begin{picture}(350,150)
\put(70,10){\includegraphics[height=150pt]{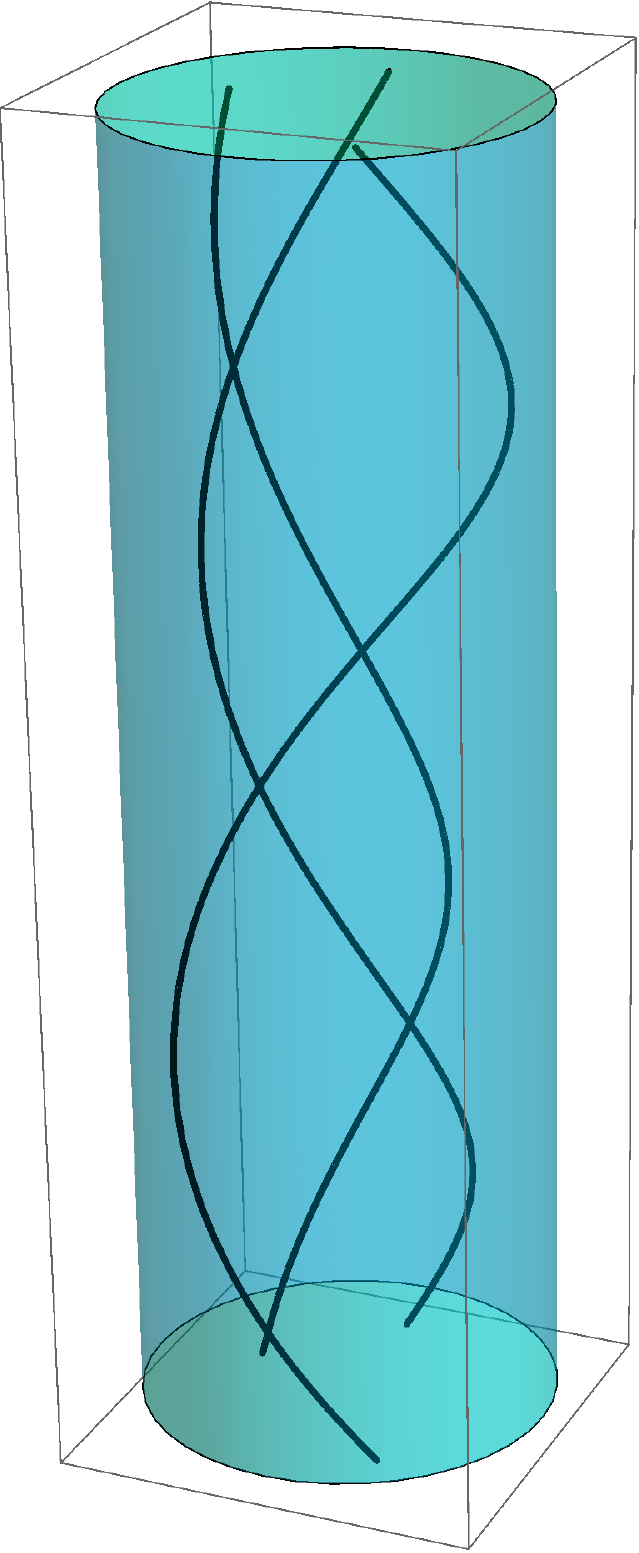}}
\put(190,40){\includegraphics[height=100pt]{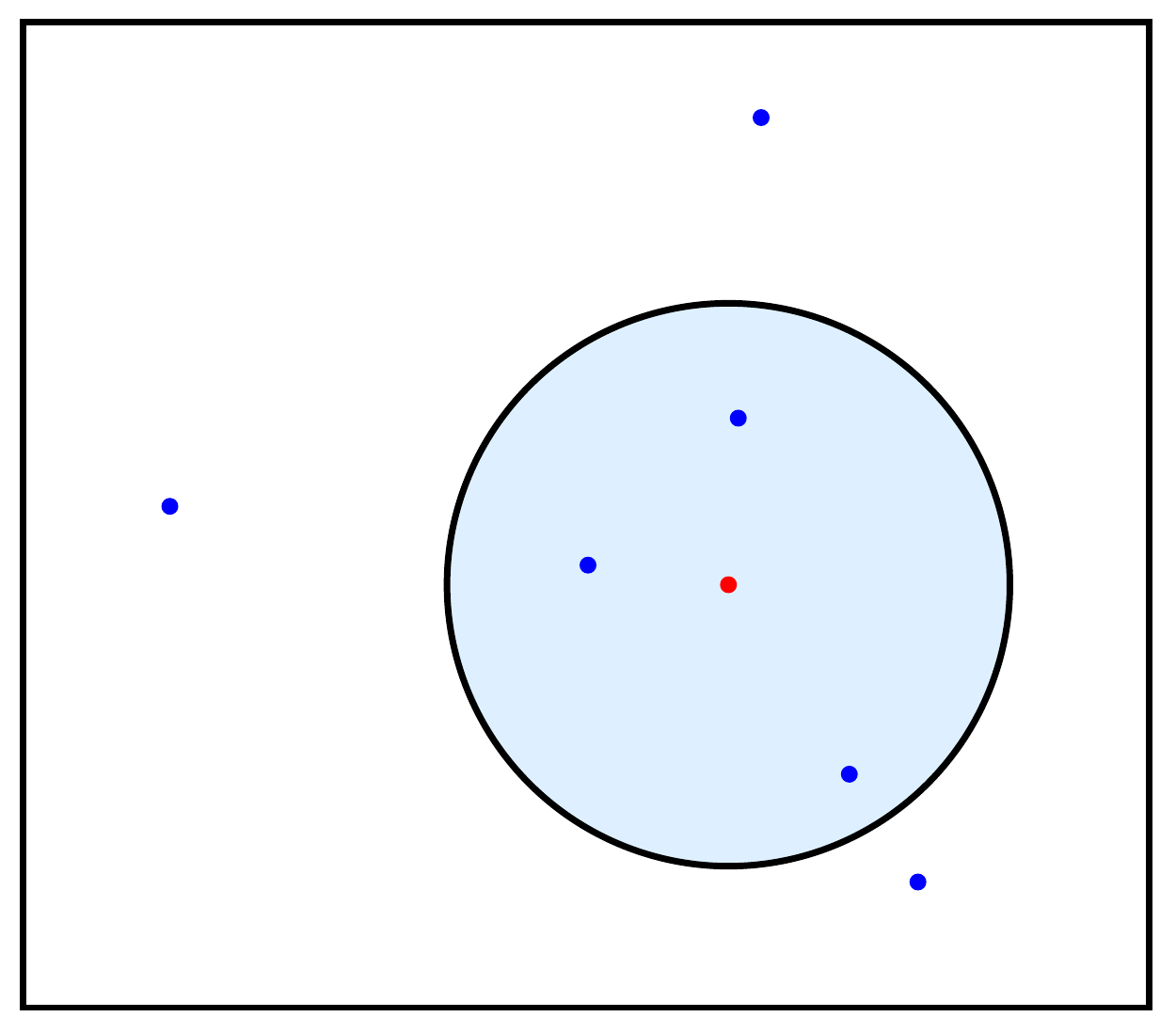}}
\put(95,0){(a)}\put(240,0){(b)}\put(260,65){$z_1$}\put(280,52){${1\over \bar z_1}$}
\end{picture}\end{center}
\caption{(a) The particle worldlines considered in this paper are helical geodesics which spiral around at constant radius in global AdS$_3$ (pictured here as a solid cylinder with time running vertically).
(b) In the dual CFT defined on the plane, our solutions represent left-right asymmetric configurations: purely holomorphic operators (blue dots) are  inserted in the points $z_i$ within the unit disk, as well as in the image points $1/\bar z_i$. A purely antiholomorphic operator (red dot) is inserted in the origin  and its image point at infinity.}\label{fig1}\end{figure}
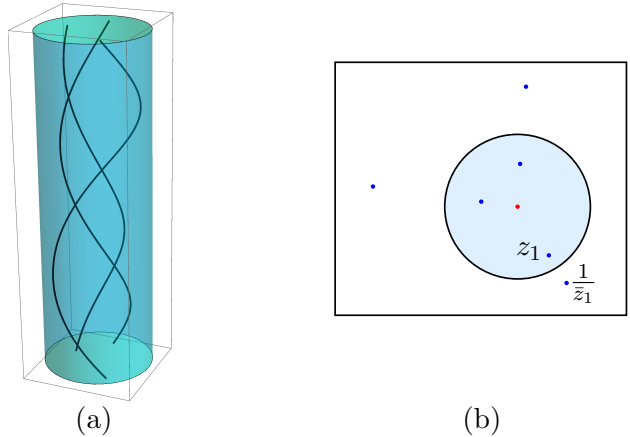

Let us now describe our setup in more detail and summarize our results. We consider 2+1-dimensional  gravity with negative cosmological constant $\L = - 1/l^2$, coupled to pointlike sources. We will focus on a particular stationary ansatz for the  metric
\be
ds^2 = l^2 \left[-  (dt - A)^2 + e^{-2\F} dz d\bar z\right]\label{metrintr}
\ee
where the function $\F$ and the one-form $A$ are defined on the base space parametrized by $z,\bar z$.
The Einstein equations reduce to the Liouville equation for $\F$ in the presence of delta-function sources:
\be
\pa_z \pa_{\bar z}\F + e^{-2\F } = 4 \p G \sum_i m_i \d^2(z- z_i, \bar z- \bar z_i).
\ee
We will see that our ansatz  describes the backreaction of point masses located at $z=z_i$, which  correspond to  helical geodesics in global AdS$_3$, see figure \ref{fig1}(a).

The coordinate $z$ can be taken to run over the unit disk, and requiring the metric to be asymptotically AdS$_3$ near the boundary $|z|=1$ imposes the boundary condition
\be
e^{2 \F} = { (1-|z|^2)^2} +\calo( (1-|z|^2)^4 )\label{ZZintr}.
\ee
In Liouville theory, this boundary condition is known as the Zamolodchikov-Zamolodchikov  \cite{Zamolodchikov:2001ah} or `pseudosphere' boundary condition (see also \cite{Menotti:2004uq},\cite{Menotti:2006gc}).
From $\F$ we can construct a holomorphic function, the Liouville stress tensor $T  = - \left( (\pa_z \F)^2 + \pa_z^2 \F\right)$. As we shall see in section \ref{secmulti}, it is of the form
\be
T =  \sum_{i=1}^N  \left({\e_i \over  (z-z_i)^2}+ {  \e_i \over  (z-1/\bar z_i)^2}+ {c_i\over z-z_i}+ {\tilde c_i\over z-1/\bar z_i}\right)\label{Tintr}
\ee
where the $\epsilon_i$ are related to the particle masses through $\e_i = 2Gm_i (1-2G m_i)$, and the $c_i, \tilde c_i$ are called accessory parameters.
When $T $ is considered as a function on the full complex plane, it has singularities at the point mass positions as well as in the image points $1/\bar z_i$.

It will turn out that  the full solution can be constructed from the knowledge of  the accessory parameters in (\ref{Tintr}), which are constrained by global considerations as follows.
It is well-known that the general solution to Liouville's equation can be written locally in terms of a holomorphic function $f(z)$ as
 \be
 e^{-2 \F} =  {|f'(z) |^2 \over (1- |f(z)|^2)^2}.\label{Phisolintr}
 \ee
When continued to the full complex plane, $f(z)$ should satisfy a `doubling trick' property  $f(z)= {1/\overline{ f (1/\bar z)}}$  in order for  $\F$ to obey the boundary condition (\ref{ZZintr}). The  function $f(z)$ and the stress tensor are related through the Schwarzian derivative
\be
S(f(z),z) \equiv {f'''\over f'} - {3\over 2} \left( { f''\over  f' } \right)^2 = 2 T (z).\label{Schwintr}
\ee
It is important to note that equation (\ref{Schwintr}) is invariant under $SL(2,\CC)$ fractional linear  transformations of  $f(z)$, i.e. $f(z) \to (a f(z) + b)/(c f(z) + d)),\ a d - b c=1$, while
 equation (\ref{Phisolintr}) is only  invariant under the subgroup of $SU(1,1)$ transformations where $c =\bar b, d = \bar a$. In particular, for generic values of the accessory parameters,
 the solution to (\ref{Schwintr}) will have a monodromy in $SL(2,\CC)$ around singular points  $z_i$ and  $1/\bar z_i$.
 The global monodromy problem we have to solve is therefore to constrain the accessory parameters $c_i, \tilde c_i$ such that the solutions to (\ref{Schwintr}) have monodromy within $SU(1,1)$, so that the Liouville field (\ref{Phisolintr}) is single-valued.

\begin{figure}\begin{center}
\includegraphics{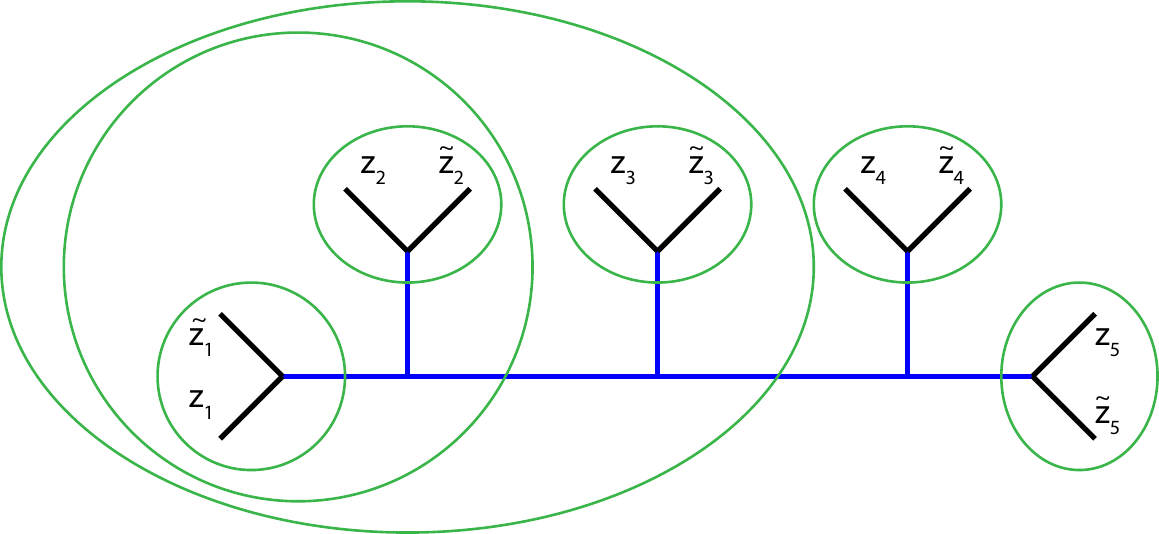}\end{center}
\caption{A diagram of a $10$-point conformal block on a sphere. The black legs represent the external primaries inserted at $z_i$ and $\tilde{z}_i \equiv \bar{z}_i^{-1}$. The blue lines correspond to the exchange of the identity family. This specific channel where the mirror pairs are fused first is the one which is relevant for the discussion of the monodromy problem on a disk. The green circles illustrate contours along which the monodromy is trivial.}
\label{fig2}\end{figure}

We will argue in section \ref{secmulti} that this  problem of determining the accessory parameters has a solution  by   relating it  to the problem of determining a
  specific  CFT conformal block at large central charge $c$. We consider the  conformal block decomposition of a $2N$-point CFT correlator on the sphere, where the operators are inserted
  in pairs of mirror points $z_i$ and $1/\bar z_i$. We focus on the channel where the operators in mirror points are fused in pairs as illustrated in figure \ref{fig2}, and furthermore specify that all the exchanged conformal families are descendants of the identity operator \footnote{This channel is related to the channel that was discussed recently in \cite{Hartman:2013mia,Banerjee:2016qca}}.
  We will then show, using the known monodromy properties of conformal blocks at large central charge \cite{Zam0},\cite{Harlow:2011ny},\cite{Hartman:2013mia},  that our accessory parameters $c_i, \tilde c_i$ can be derived from the knowledge of this particular conformal block. The problem of finding the gravity solution is then reduced to integrating (\ref{Schwintr}), which is well-known to reduce to solving an auxiliary linear ordinary differential equation.

One peculiar feature of our setup is that   our particles move on geodesics in Lorentzian signature, while the connection between geodesics and conformal blocks is best understood for Euclidean geodesics.
Under analytic continuation, our Lorentzian geodesics generically continue to {\em complexified} geodesics in Euclidean AdS$_3$. We will argue in section \ref{sechol}, from
considering the holographic stress tensor, that our solutions
 represent {\em left-right asymmetric} configurations in the dual CFT defined on the plane, with purely holomorphic operators  inserted in the points $z_i$ within the unit disk, as well as in the image points $1/\bar z_i$ (see figure \ref{fig1}(b)).

 Finally, in section \ref{secabel}, we will  briefly comment on multi-centered solutions for which the metric is static rather than stationary, and give a new perspective on the observed phenomenon  that such solutions always seem to have
additional unphysical singularities \cite{Clement:1994qb},\cite{Coussaert:1994if},\cite{Mansson:2000sj}. We will show that a static ansatz restricts the monodromy group discussed  above  to be abelian. It is known in the theory of Fuchsian differential equations (see e.g. \cite{Yoshida}), that placing such  strong restrictions on the monodromy group does not lead to a solution unless one introduces so-called spurious singularities in the differential equation (\ref{Schwintr}). These are precisely the unphysical singularities
found in \cite{Clement:1994qb},\cite{Coussaert:1994if},\cite{Mansson:2000sj}.

\section{Ansatz for helical multi-particle solutions}\label{secansatz}
In this work we will revisit the construction of solutions to Einstein gravity  with negative cosmological constant $\L$ in 2+1 dimensions in the presence of point particle sources. The action, up to boundary terms, is
\bea
S&=& {1\over 16 \p G} \int_\calm d^3 x\sqrt{-g} \left( {R } -2 \L \right) + S_{\rm source} \\
S_{\rm source} &=& -  \sum_i {m_i } \int_{w_i} ds_i\label{einstact}
\eea
where $w_i$ are the timelike worldlines of the particle sources and $m_i$ are the masses of the particles. The sources are required to move on geodesics of the backreacted metric,  which is necessary for the stress-energy  tensor computed from $S_{\rm source}$  to be conserved. The equations of   motion following from (\ref{einstact}) are
\be
G^{\m\n} +\L g^{\m\n}= 8 \p G \sum_i m_i \int d\t_i {\d^3(x- x_i(\t_i)) \over \sqrt{|g|} } {\dot x_i^\m \dot x_i^\n \over \sqrt{|g_{\r\s} \dot x_i^\r \dot x_i^\s|}} .\label{Einst}
\ee
Since in 2+1 dimensions the Riemann tensor is completely determined by the Einstein tensor, the metric is locally $AdS$ outside of the sources, whose effect is however imprinted on the global structure of the manifold  in the form of deficit angles as we will review below.  Even though naively there are no attractive forces between particles, these global effects can lead to nontrivial dynamics, e.g. after backreaction the geodesics of two particles can merge to form a single BTZ black hole \cite{Matschull:1998rv}.

In this work we will mostly be concerned with the backreaction of   particles moving on a specific class of helical geodesics for which the problem simplifies greatly.  These can be viewed as the closest analogy in AdS$_3$ of the backreaction of particles along static geodesics in Minkowski space.
First of all, we will limit ourselves to solutions which preserve a timelike Killing vector. The  particle worldlines we consider are
then integral curves of  this Killing vector. Choosing adapted  coordinates $(t,z ,\bar z)$ such that the Killing vector is $\pa/ \pa t$, the worldlines  are the curves of constant $z,\bar z$. The most general form of the metric is
\be
ds^2 = - N^2 (dt - A)^2 + e^{-2\F} dz d\bar z\label{genKV}
\ee
where $N, \F$ and the one-form $A$ are defined on the 2-dimensional base parametrized by $(z, \bar z)$. Note that we have chosen  a conformal gauge for the 2-dimension spatial base metric. The Einstein equations (\ref{Einst}) lead to the following system of equations
\bea
\pa_z \left( \pa_z N e^{2\F}\right) &=&0\label{Einst1}\\
\pa_z \left(  N^3  e^{2\F} F_{z\bar z} \right) &=&0\\
2 \pa_z\pa_{\bar z} N + N \left( \L e^{-2\F}  - N^2  e^{2\F} (F_{z\bar z})^2\right) &=&0\\
4 \pa_z\pa_{\bar z}\F -\L e^{-2\F} - 3 N^2  e^{2\F} (F_{z\bar z})^2 &=& 16 \p G \sum_i m_i \d^2(z- z_i, \bar z- \bar z_i)\label{Einst4}
\eea
where $F=dA$ is the field strength of $A$. We note that the first two equations are complex and constitute two real equations each.

As is evident from the first equation, the solutions will be qualitatively quite different depending on whether $N$ is constant or not, and we will mostly
focus on the latter case.
  Let's first consider this case without sources. A constant $N$ implies that there is no gravitational potential with respect to $\pa/\pa t$, or in other words that every curve of constant $z, \bar z$ is a geodesic. Upon introducing particle sources,
the equations imply that $N$ remains constant while $F$ and $\F$ obey\footnote{The sign in front of $F_{z\bar z}$ is a matter of convention, since the Einstein equations (\ref{Einst1}-\ref{Einst4}) are invariant under $F_{z\bar z} \to - F_{z\bar z}$.}
\bea
F_{z \bar z} &=&- { i \sqrt{-\L} e^{-2 \F} \over N }\label{Feq1}\\
\pa_z \pa_{\bar z}\F - \L e^{-2\F } &=& 4\p G \sum_i m_i \d^2(z- z_i, \bar z- \bar z_i).  \label{Liouv1}
\eea
In the flat space ($\L =0$) case, $F=0$ and the metric is static while $\F$ must satisfy  Poisson's equation with point-like sources.  This is the situation discussed in the classical work \cite{Deser:1983tn}. In the $AdS$ case
of interest ($\L <0$)  $F$ is nonvanishing and the metric  is necessarily
stationary rather than static\footnote{Note that $F_{z\bar z}$ is imaginary so that (\ref{Feq1}) makes sense for negative $\L$.}. The field $\F$ now satisfies a Liouville equation with  delta-function sources.  We note also from (\ref{Feq1}) that in (Lorentzian) de Sitter space ($\L >0$),  solutions  with constant $N$ would not be possible.
When, on the other hand, $N$ is not constant, the sources do backreact on $N$ and the equations become more complicated.
For  a static ansatz (i.e. $F=0$), this was discussed in \cite{Deser:1983nh}; we will comment only briefly on this case from our point of view in section \ref{secabel}.

Returning to the case of negative $\L$ and constant $N$, we can set $N$ to one without loss of generality. We will also set $\L = - 1/l^2$ and for later convenience introduce dimensionless coordinates rescaled by $l$, such that our ansatz (\ref{genKV})  and equations of motion (\ref{Feq1},\ref{Liouv1}) become
\bea
ds^2 &=& l^2 \left[-  (dt - A)^2 + e^{-2\F} dz d\bar z\right] \label{ansatz}\\
F_{z \bar z} &=& {- i  e^{-2 \F} }\label{Feq}\\
\pa_z \pa_{\bar z}\F + e^{-2\F } &=&  4\p G \sum_i m_i \d^2(z- z_i, \bar z- \bar z_i).  \label{Liouv}
\eea
The ansatz (\ref{ansatz}) also  arises naturally in  studies of BPS solutions in 3D supergravity \cite{Levi:2009az}.
We note that (\ref{ansatz}) is invariant under conformal transformations on the base with the Liouville field $\F$ transforming in the standard way,  provided we also shift the time coordinate:
\bea
z &\to& \tilde z(z) \label{conftransf1} \\
\F(z, \bar z) &\to& \tilde \F (\tilde z, \bar{\tilde z}) = \F (z, \bar z) + \ln |\pa_z \tilde z |\\
t &\to & \tilde t = t + \half \Im m \ln \pa_z \tilde z .\label{conftransf}
\eea
From the Liouville field $\F$, we can construct a holomorphic quantity, namely the Liouville stress tensor
\be
 T  (z) = - \left( (\pa_z \F)^2 + \pa_z^2 \F\right). \label{TLiouv}
 \ee
This object will play an important role in what follows; indeed it will turn out the full solution can be reconstructed from  knowledge of $T $.
Under conformal transformations (\ref{conftransf1}-\ref{conftransf}), $T $ transforms as $T (z) \to \tilde T  (\tilde z)$ with
\be
(g')^2 \tilde T (\tilde z)  = T (z) -\half  S(\tilde{z},z)
\ee
where $S(f(z),z)$ is the Schwarzian derivative defined in (\ref{Schwintr}).

To get some intuition for the particle configurations described by our ansatz, it is useful to rewrite the  global AdS metric
\be
ds^2 = l^2 \left[-\cosh^2 \r d\t^2 +  d\r^2 + \sinh^2 \r d\f^2\right]\label{globalads}
\ee
in the form (\ref{ansatz}).  One finds that the coordinate transformation
\bea
z &=& e^{i (\t+\f)} \tanh \r \\
t &=& \t\label{toglobal}
\eea
brings (\ref{globalads}) into the form (\ref{ansatz}), with
\be
e^{2 \F} = { (1-|z|^2)^2}\label{Phiglobal}
\ee
and $A$ related to $\F$ by (\ref{Feq}). The particle sources at constant $z, \bar z$ correspond to constant radius $\r$ and constant $x_+ = \t + \f$ in global AdS$_3$. Therefore we are constructing  the backreaction of
particles on helical geodesics which spiral  around at constant radius  in global AdS$_3$ as shown in figure \ref{fig1}(a).

Let us also give a heuristic derivation of the dual interpretation of these helical solutions, which will be confirmed by explicit calculation in section \ref{sechol}.
 The interpretation of bulk geodesics in the  dual CFT
is most   clearly understood in the case of geodesics in Euclidean AdS. These always have two endpoints on the boundary, and correspond in the dual CFT to  insertions of scalar primary operators at the endpoints \cite{Fitzpatrick:2014vua}.
The geodesics we are considering are, however, timelike geodesics in Lorentzian AdS which (in general) don't admit a natural continuation to Euclidean signature. Indeed, our  geodesics are at constant $z = z_0$ which in global coordinates reads
\be
\r = \r_0 ,\qquad  \t + \f = x_{+,0}.
\ee
If we continue to Euclidean signature by continuing the global time $\t \to - i \t_E$,  our geodesics obviously become complex, with the exception of the geodesic at $z_0=0$, i.e. in the center $\r_0=0$ of global AdS$_3$.
This latter geodesic has endpoints at $\t_E = \pm \infty$ or, after conformally mapping  to the complex plane with coordinate $u$, at $u=0$ and $u=\infty$. Hence in the dual CFT, scalar primaries are inserted at $u=0$ and $u=\infty$. To interpret our other worldlines in $z_0\neq 0$, we will use the fact  that they are simply related to the one in $z_0=0$ by symmetry. Indeed, we can represent  points $(t,z, \bar z)$ in Lorentzian
AdS as $SU(1,1)$ group elements
\be
g (t,z,\bar z) = {1\over \sqrt{ 1- |z|^2}} \left( \begin{array}{cc} 1 & z\\ \bar z & 1\end{array}\right) \cdot \left( \begin{array}{cc} e^{i t} & 0\\ 0 &  e^{-i t}\end{array}\right).
\ee
The worldline of a particle at $z=z_0$ is described by $g(t, z_0,\bar z_0)$ and is obviously related to the one at $z=0$ by left multiplication by a constant group element:
\be
   g (t,z_0,\bar z_0) = {1\over \sqrt{ 1- |z_0|^2}} \left( \begin{array}{cc} 1 & z_0\\ \bar z_0 & 1\end{array}\right) \cdot g(t,0,0).
   \ee
This will act in the dual CFT by the corresponding purely holomorphic fractional linear transformation
\be
u \to { u + z_0 \over \bar z_0 u +1}, \qquad \bar u \to \bar u.
\ee
This maps  the scalar operators in $u=0$ and $u =\infty$  to a configuration with purely holomorphic  operators in $u = z_0$ and $u = 1/\bar z_0$ respectively, and purely anti-holomorphic operators in the origin and at infinity, as illustrated in figure \ref{fig1}(b).

After these preliminaries we turn to the problem of constructing solutions. The equation  (\ref{Feq}) is readily solved: given a solution to the Liouville equation (\ref{Liouv}), the solution for the gauge potential $A$ is, up to an exact form which can be absorbed in a redefiniton of $t$,
\be
A = \Im  \left ( \pa_z \F   dz + b d \l \right) \label{Asol}
\ee
where
\be
\l = -2 G \sum_i m_i \ln {z-z_i \over 1- \bar z_i z}.\label{lambdadef}
\ee
The large gauge transformation involving the multivalued function $\l (z)$ ensures that $A$ is free of Dirac string singularities, as required by the equations of motion (\ref{Einst}).
Note that we have introduced a constant $b$ which should be taken to be 1 in order to describe our current setup. As we will argue in section \ref{sechol}, setting $b$ to zero gives solutions to equations with different delta-function source terms which, which are appropriate to describe extremal spinning particles.

We are hence left with the Liouville  equation (\ref{Liouv}),  which is to be solved on a manifold with boundary. Therefore we should first  discuss the boundary condition to be imposed on the Liouville field.
By a conformal mapping,
we can assume that the coordinates $z,\bar z$ in our ansatz (\ref{ansatz}) range over the unit disk $|z|\leq 1$ as was the case  for the global AdS$_3$ solution (\ref{toglobal}). Since the Liouville field describing global AdS$_3$ is given by (\ref{Phiglobal}),
 we will impose the following boundary condition\footnote{We  show that the $\calo( (1-|z|^2)^3 )$ term is absent  in Appendix \ref{appas}.}  for $|z|\to 1$:
\be
e^{2 \F} = { (1-|z|^2)^2} +\calo( (1-|z|^2)^4 ). \label{Liouvbc}
\ee
With this boundary condition, the metric $e^{-2 \F} dz d\bar z$ on the base manifold asymptotically approaches the hyperbolic metric on the Poincar\'e disk (also called  pseudosphere). This boundary condition
 was studied by Zamolodchikov and  Zamolodchikov
\cite{Zamolodchikov:2001ah} and we will refer to it as  the `ZZ boundary condition'. Since this boundary condition preserves conformal invariance, it leads to the standard reality condition on the stress tensor $T $ which
 expresses that   there is no energy or momentum flow through
the boundary. For $T $ defined on the unit disk, this condition reads
 \be
 \left. \left( {z^2} T (z)\right)\right \rvert_{|z|=1} \in \RR.\label{realTL}
 \ee
 We will explicitly derive this property from the boundary condition (\ref{Liouvbc}) in section \ref{sechol}  below. Even though we will work in the coordinate $z$, defined on the unit disk, for the remainder of this work, it may be useful to work out what (\ref{realTL}) would look like when working on the upper half plane. Conformally transforming
 $T  (z)$ to $\tilde T  (u)$ defined on the upper half plane  using the Cayley map $z= {u-i \over u+i}$,
 (\ref{realTL}) becomes the familiar condition \cite{Cardy:1984bb}
that $\tilde T  (u)$ is real when $u$ is real.

\section{Multi-centered solutions from conformal blocks}\label{secmulti}

In the previous section we showed that the multi-centered solutions of interest are constructed from   solutions to  the Liouville equation on the unit disk  $\rm{D}$, $|z|<1$, in the presence of  delta-function sources
\be
\pa_z \pa_{\bar z}\F + e^{-2\F } = \half \sum \d_i \d^{(2)} ( z- z_i,\bar z - \bar z_i)\, ,
\label{Liouvsource}
\ee
with prescribed asymptotic boundary conditions (\ref{Liouvbc}) for $|z| \to 1$. We have introduced the dimensionless quantities $\d_i \equiv 8 \p G m_i$.

If we can neglect the Liouville potential term compared to the kinetic term near the sources, the field $\F$ behaves near $z = z_i$
as
\be \F \sim {\d_i \over  2 \p}\ln |z-z_i|.\label{Phinearsing} \ee
Substituting in the action for the Liouville field shows that neglecting the potential was justified for $\d_i < 2\p$. This bound has a simple geometric meaning:
substituting (\ref{Phinearsing}) into the metric (\ref{ansatz}), we see that the  metric $e^{-2 \F} dz d\bar z$ on the base manifold has a conical singularity with deficit angle $\d_i$, and the above bound
says  that the deficit angles are bounded from above by $2\p$, or in other words that the opening angles are nonnegative. We will mostly restrict attention to positive masses $m_i$ and hence positive $\d_i$, although negative $\d_i$,  corresponding to   excess angles, will make an appearance in section \ref{secabel}.

Note that on the disk there is no further restriction on the deficit angles $\delta_i$. This is unlike the situation in the case of Liouville equation on the sphere where the Gauss-Bonnet theorem implies an inequality for the total sum of deficit angles (see e.g. \cite{Seiberg:1990eb}). The usual argument involves the assumption of compactness of $S^2$ and in the case of the disk the presence of the asymptotic boundary of $D$ where $\Phi$ diverges invalidates this argument. Note that, for the same reason, the standard arguments for the existence and uniqueness of solutions to the Liouville equation on compact manifolds don't generalize in a straightforward manner to the case at hand.

In order to solve the equation, we will proceed in several steps which are summarized as follows:
\begin{itemize}
\item Assuming that there is a solution of the Liouville equation (\ref{Liouvsource}), it determines a meromorphic stress-energy tensor $T (z)$ (\ref{TLiouv}) defined in the disk. $T (z)$ can have at most double poles at the location of the sources.
\item The boundary conditions (\ref{realTL}) allow us to analytically continue $T (z)$ to the whole complex plane. The resulting meromorphic function $T (z)$ will have at most double poles at the locations of the sources and their mirror images. The coefficients of the double poles are determined by the deficit angles $\delta_i$, while the coefficients of simple poles (called accessory parameters) are at this stage unknown  and we have to determine them later.
\item The solution of Liouville equation (\ref{Liouvsource}) can be reconstructed from $T (z)$ if we can solve a (holomorphic) Schr\"odinger equation in potential $-T (z)$. There is a $SL(2,\mathbb{C})$ freedom in the choice of two linearly independent solutions of this equation and we will see that the boundary condition (\ref{Liouvbc}) will reduce this freedom to $SU(1,1)$. The remaining $SU(1,1)$ freedom in turn does not change the Liouville solution $\Phi(z,\bar{z})$, so knowing $T (z)$ we can reconstruct a unique $\Phi(z,\bar{z})$.
\item The remaining problem is to determine the accessory parameters in $T (z)$. It turns out that the the two solutions of associated Schr\"odinger equation have a non-trivial $SL(2,\mathbb{C})$ monodromy around each source point and that only in the case that these monodromy matrices are in $SU(1,1)$ the resulting Liouville field will be single-valued. We thus have to tune the accessory parameters in such a way that the singularities have $SU(1,1)$ monodromy instead of the general $SL(2,\mathbb{C})$ monodromy and the Liouville field is single-valued.
\item A related monodromy problem has been studied in the context of classical conformal blocks. There, one knows the coefficients of the double poles in terms of conformal dimensions of external primaries, while the dimensions of exchanged primaries enter through the conjugacy class of the monodromy matrix around the singular points. The accessory parameters are thus again determined in terms of monodromies, but instead of fixing the subgroup where the monodromy matrix lies we have to fix the conjugacy class.
\item Returning to our monodromy problem, by computing the monodromy around the insertion point and its mirror, we find that they are in fact each others inverses, so the monodromy around each mirror pair is the identity. This is also true for monodromies around any number of mirror pairs. Comparing this observation with the monodromy problem for classical conformal blocks, we find that our monodromy problem is
    equivalent to a classical conformal block problem where the external primaries are fused first in mirror pairs and all the exchanged internal primary fields are the identity. In this way, the multi-centred gravity solution can be constructed, in principle, from the knowledge of a classical conformal block by an integration of second-order linear ODE.
\end{itemize}
As an example of this procedure, we conclude this section by showing a computation of solution with two sources in the disk, where one of them is light and is treated as a small perturbation of one-source solution.

\subsection{Properties of the Liouville stress tensor}

\paragraph{Pole structure} Each solution $\Phi$ of the Liouville equation (\ref{Liouvsource}) on the disk determines a meromorphic stress tensor
\begin{equation}
T  (z) = -\left( (\pa_z \F)^2 + \pa_z^2 \F\right).
\end{equation}
Although the exact form depends on the particular solution, from (\ref{Phinearsing}) we find that the singularity structure is fixed by the sources (i.e. by $\delta_i$) as
\be
T  (z) = \sum_{i =1}^N {\e_i \over (z-z_i)^2} + \ldots
\ee
where $\e_i =  { \d_i \over 4 \p} \left( 1-  { \d_i \over 4 \p} \right)$.
The ellipses denote lower order terms, i.e.  first order poles and a regular part. As we can see $T  (z)$ is meromorphic on the unit disk with at most second-order poles at positions of sources with coefficients fixed solely by $\delta_i$.

\paragraph{Reality condition and and Schwarz reflection principle}
Since the boundary value of $T(z)$ must satisfy the reality condition \eqref{realTL}, we observe that $z^2 T $ is a meromorphic map from the unit disk such that the unit circle is mapped to the real line. The  Schwarz reflection principle applied to functions which map the unit  circle to real line (see e.g. \cite{Ahlfors}) then implies that $T $ can be analytically continued to a meromorphic function in the whole complex plane. The resulting function on the plane is defined as
\be
z^2 T (z) = \begin{cases} z^2 T (z), & |z| \leq 1, \\ {1 \over z^2} \bar T  (1/z), & |z|>1. \end{cases}
\label{reflTLdef}
\ee
and using the same name $T (z)$ for the function defined in the whole complex plane, by construction it has the reflection property
\be
T (z) = {1 \over z^4} \bar T  (1/z).
\label{reflTL}
\ee

\paragraph{Constraints on the accessory parameters}
Now that we have continued $T $ to the whole complex plane, let us now see what are the constraints on the form of $T $ following from (\ref{reflTL}). $T $ has poles both in the $z_i$ and in their image points  ${\bar z_i}^{-1}$. Assuming first for simplicity that none of the sources is at the origin (and therefore that there is no image source at infinity), $T $ must be of the form\footnote{The assumed structure of poles would allow for addition of an arbitrary polynomial - but this would lead to unwanted higher order poles at infinity which would violate the reflection property (\ref{reflTL}).}
\be
T =  \sum_{i=1}^N  \left({\epsilon_i \over  (z-z_i)^2}+ { \tilde \epsilon_i \over  (z-1/\bar z_i)^2}+ {c_i\over z-z_i}+ {\tilde c_i\over z-1/\bar z_i}\right).\label{accesspar1}
\ee
The point particle deficit angles $\d_i < 2\p$ lead to  values of $\epsilon_i = {\d_i \over 4 \p}\left( 1- {\d_i\over 4\p} \right)$ which are smaller than $1/4$, but we will also briefly comment on the physical meaning of the case  $\epsilon \geq 1/4$ below.
The residues $c_i, \tilde c_i$ of the single poles  are called accessory parameters and they are not determined in terms of $\delta_i$ but will instead be determined later by solving a monodromy problem.

Substituting (\ref{accesspar1}) into (\ref{reflTL}) and demanding the equality of single and double pole terms near $z = z_i$ leads to
\bea
\epsilon_i - \bar{\tilde \epsilon}_i &=&0\label{reflcond1}\\
2 \epsilon_i + c_i z_i + {\bar{\tilde c}_i \over z_i} &=& 0.\label{reflcond2}
\eea
Substituting these relations into (\ref{reflTL}) one finds only two further conditions from the requirement of the regularity of $T $ at origin:
\bea
\sum_{i=1}^N (c_i + \tilde c_i) &=&0\label{reflcond3}\\
\sum_{i=1}^N \Im m  \left(c_i z_i  - {\bar{\tilde c}_i\over z_i} \right)&=&0\label{reflcond4}
\eea
Until now, we assumed that none of the sources are at the origin, and hence none of the image charges to be at infinity. If we allow for a double pole at origin, instead of (\ref{accesspar1}) we have
 \be
T = {\epsilon_0 \over z^2} + {c_0\over z} +\sum_{i=1}^{N-1}  \left({\epsilon_i \over  (z-z_i)^2}+ { \tilde \epsilon_i \over  (z-1/\bar z_i)^2}+ {c_i\over z-z_i}+ {\tilde c_i\over z-1/\bar z_i}\right) \label{accesspar2}
\ee
($N$ still denotes the number of insertions inside of the unit disk). The conditions (\ref{reflcond1}-\ref{reflcond2}) for $i \geq 1$ remain the same while (\ref{reflcond3}-\ref{reflcond4}) generalize to
\bea
c_0 + \sum_{i=1}^{N-1} (c_i + \tilde c_i) &=&0\label{0reflcond3}\\
\Im m \left( 2 \e_0 + \sum_{i=1}^{N-1}  \left(c_i z_i  - {\bar{\tilde c}_i\over z_i} \right)\right)&=&0\label{0reflcond4}
\eea
In both cases, if we fix the $\epsilon_i$, the $\tilde\epsilon_i$ are fixed through (\ref{reflcond1}). Eqs. (\ref{reflcond2}-\ref{reflcond4}) resp. (\ref{reflcond2},\ref{0reflcond3}-\ref{0reflcond4}) further reduce the number of accessory parameters $c_i, \tilde c_i$ to $2N-3$ real parameters where $N$ always denotes the number of insertion points inside of the unit disk. This should be compared to the case of the Riemann sphere with $N$ insertion points where the number of undetermined real accessory parameteres is $2N-6$. The number of relations between the accessory parameters reflects nicely the number of real isometries of the space which is $6$ in the case of the round Riemann sphere and $3$ for the hyperbolic unit disk.

\subsection{Associated ODE problem}

It is matter of simple calculation to show that $e^\F$ satisfies the following differential equations
\be
\left( \pa_z^2 + T (z) \right) e^{\F(z, \bar z)} = 0,
\qquad
\left( \pa_{\bar z}^2 + \bar T (\bar z) \right)e^{\F(z, \bar z)} =0 \,.
\ee
From the general theory of linear $2^{nd}$ order ODE's we know that every solution can be written as a linear combination of two linearly independent solutions. This allows us to write $\F$ in the factorized form
\be
e^\F = \psi_1 (z) \tilde \psi_1 (\bar z) +  \psi_2 (z) \tilde \psi_2 (\bar z)\label{LiouvFuchs}
\ee
with $\psi_{1,2}$ and $\tilde \psi_{1,2}$ independent holomorphic and anti-holomorphic solutions of the ODE and its complex conjugate respectively
\be
\left( \pa_z^2 + T (z) \right)\psi (z) =0 \,,\qquad
\left( \pa_{\bar z}^2 + \bar T (\bar z) \right) \tilde \psi (\bar z) =0 \,.\label{Fuchs1}
\ee

It will be convenient to arrange the solutions in to a column vector designated as $\Psi (z) = (\psi_1,\psi_2)^T$. The Wronskian
$W\equiv \psi_1' \psi_2 - \psi_1 \psi_2'$ is constant and by an appropriate rescaling $\Psi \to \l \Psi $ and $ \tilde \Psi \to \tilde \Psi/\l$, we can normalize $W=1$.
It should be noted that $\Psi$ and $\tilde{\Psi}$ are not necessarily each other's complex conjugates,  even though they solve  conjugate equations. In general,  $\bar{\psi}_i$ will be linear combinations of $\tilde{\psi}_i$,
\be
\tilde \Psi = \L^T \bar \Psi
\ee
for some constant matrix $\L \in GL(2,\CC)$, so that
\be
e^{\F(z, \bar z)} =  \Psi^\dagger (\bar z)  \L \Psi (z).
\ee
Substituting this in the Liouville equation (and using a Fierz-like rearrangement formula as well as the Wronskian condition) yields a restriction on $\Lambda$
\be
\det \L = -1.
\ee
In order to obtain a real, positive definite metric, we need $e^{-2\F}$ to be real and positive, or equivalently $e^\F$ to be real. This imposes one further condition
\be
\L^\dagger = \L.
\ee
By making a change of basis in the vector space of solutions to (\ref{Fuchs1}), $\Psi \to M \Psi$, with $M \in SL(2,\CC )$, we can bring $\L$ into a canonical form. The one which we will use here is
\be
\L = \begin{pmatrix} 1 & 0 \\ 0 & -1 \end{pmatrix} \equiv \sigma_3. \label{Lambdas3}
\ee
The Liouville  solution then   takes the form
\be
e^{-2 \F} = (\psi_1 \bar \psi_1 - \psi_2 \bar \psi_2)^{-2}. \label{ourframe}
\ee
With our choice of canonical form  of $\L = \sigma_3$ , the Liouville solution $\F$ is invariant  under transformations $\Psi \to M\Psi$ with $M^\dagger \s_3 M = \s_3$.
  Such transformations form the group  $SU(1,1)$, and will be discussed in more detail at the end of this section.

It's useful to introduce a function $f(z)$ through
\be
f =  {\psi_1 \over \psi_2}.\label{deff}
\ee
The knowledge of $f$ implies knowledge of $\psi_{1,2}$,
since (\ref{deff}) can be inverted using the Wronskian condition as follows
\be
 \psi_1 = {f \over \sqrt{ f'}},\qquad
\psi_2 = {1 \over \sqrt{ f'}}.\label{psiitof}
\ee
The solution (\ref{ourframe})  for the Liouville field becomes
\be
e^{-2 \F} =  {|f'|^2 \over (1- |f|^2)^2}.\label{ourframef}
\ee
Substituting in the metric $e^{-2 \F} dzd\bar z$ on the 2D base in (\ref{ansatz}) shows that it is the pull-back of the constant negative curvature metric on unit disk with respect to this $f$\footnote{Had we chosen another canonical form for $\L$ than the one  in (\ref{Lambdas3}), we would have obtained the pullback of a conformally related metric,
e.g. the choice $\L =  \s_2$ would result in the pullback of the  constant negative curvature metric on the upper half plane.}.
The Liouville stress tensor is proportional to the Schwarzian derivative of $f$
\be
2T (z)  = {f'''\over f'} - {3\over 2} \left( { f''\over  f' } \right)^2\equiv  S(f(z),z).\label{schweq}
\ee
\paragraph{Boundary condition}
We can express the ZZ boundary condition (\ref{Liouvbc}) in terms of $\Psi$ or $f$, leading to
\bea
|\psi_1 (z)|^2 = |\psi_2 (z)|^2 \qquad && {\rm for\ } |z|=1.\label{ZZpsi}\\
|f (z)|^2 = 1 \qquad  && {\rm for\ } |z|=1 .\label{ZZf}
\eea
One can check that this indeed  leads to the asymptotic behaviour (\ref{Liouvbc}), and it will follow from the near-boundary analysis that the most general Liouville solution with ZZ boundary conditions can be described by functions $\Psi, f$ satisfying these properties. The condition (\ref{ZZf}) states that  $f$  maps the boundary circle to a unit circle. Therefore we can extend $f(z)$ to the full complex plane using the Schwarz reflection principle (for maps  which map the unit circle to the unit circle), and the resulting function satisfies
\be
f(z)= {1\over \overline{ f (1/\bar z)}}= {1\over \overline{ f} (1/ z)}.\label{reflf}
\ee
In terms of $\Psi$, the reflection property is
\begin{equation}
\Psi(z) = z \sigma_1 \bar{\Psi}\left(\frac{1}{z}\right).\label{reflPsi}
\end{equation}

It will be important later on that, starting from  a solution of the differential equation (\ref{Fuchs1}) or (\ref{schweq}), we can always reach a solution satisfying the reflection property (\ref{reflPsi}) or (\ref{reflf})
by making a suitable $SL(2,\CC)$ transformation. This is shown in appendix \ref{appA}. As already mentioned, the Liouville field is invariant under $SU(1,1)$ transformations,
 which act on $\Psi$ by matrix multiplication and on $f$ as fractional linear transformations as follows
\bea
\Psi & \to & \left(\begin{array}{cc} a& b\\\bar b & \bar a \end{array}\right) \Psi, \qquad |a|^2-|b|^2=1 \\
 f &\to& { a f + b \over \bar b f + \bar a}.
\eea
One can also show that these transformations also preserve the reflection properties (\ref{reflf}) and (\ref{reflPsi}) of $f$ and $\Psi$.

\subsection{The accessory parameter problem}

At this point, it only remains to determine the accessory parameters  in $T $ and we will do this by studying the monodromy of $\Psi$ and $f$ when encircling the particle sources. For generic values of the accessory parameters,  the monodromy of a solution $\Psi$ to (\ref{Fuchs1}) around the singular points will be in $SL(2,\CC)$, i.e. after encircling a singular point $z_i$ the solutions of the ODE transform as
\begin{equation}
\Psi \to M_i \Psi, \qquad M_i = \begin{pmatrix} a_i & b_i \\ c_i & d_i \end{pmatrix} \in \rm{SL}(2,\mathbb{C})
\end{equation}
while $f$ undergoes a linear fractional transformation
\be
f \to \frac{a_if+b_i}{c_if+d_i} \,.
\ee
The Liouville field $e^{\F}$  is not a single-valued function unless $M_i$ is actually in $SU(1,1)$. Therefore, in order to have a single-value solution of Liouville equation we must make sure that all $M_i$ are elements of $SU(1,1)$ by adjusting the values of accessory parameters.

\paragraph{Conjugacy classes of monodromy matrices}\label{secmon}
The conjugacy class of the monodromy matrix $M$ is determined by the coefficient $\e_i$ of the double pole in $T $ at the singular point $z_i$. From the behaviour of solutions of (\ref{Fuchs1}) near a singular point one easily checks that the exponents of the differential equation associated to $z_i$ are \footnote{Recall that for a differential equation with regular singular point $z_0$ the exponents associated to this singular point are complex numbers $\alpha$ such that in the neighbourhood of $z_0$ there is a solution whose leading order behaviour is $\sim(z-z_0)^\alpha$.}
\begin{equation}
\frac{1}{2} \pm \frac{1}{2} \sqrt{1-4\epsilon_i} = \begin{cases} \delta_i/4\pi \\ 1-\delta_i/4\pi \end{cases}
\end{equation}
so the trace of the monodromy matrix is
\be
\tr M_i = -2\cos \p \sqrt{1-4\epsilon_i} = 2\cos\left(\frac{\delta_i}{2}\right).
\ee
For $M_i \in SU(1,1)$ this must be real, so we see that unless $\delta_i$ is real or pure imaginary we cannot expect a solution of our monodromy problem. Although our main interest is in the case $0<\epsilon_i < 1/4$  we will give here a brief overview (see also \cite{Seiberg:1990eb},\cite{Martinec:1998wm})  of the other possibilities for the coefficients of the double poles in $T $ and their physical interpretation:
\begin{itemize}
\item {\bf Elliptic singularity:} For $\e_i<1/4$, as is the case for the solutions of interest, the monodromy matrix belongs to an elliptic conjugacy class of $SU(1,1)$, with purely imaginary non-degenerate eigenvalues.
\item {\bf Spurious singularity:} For the special values $\e _i= (1- n^2)/4 , n\in \NN / \{0\}$ in the case above, the monodromy  actually becomes trivial. Such a singularity in $T $ is called spurious \cite{Yoshida}. For $n=1$ we recover global $AdS$ and the metric is smooth, but for $n>1$ there is a delta-function curvature singularity corresponding to a negative deficit angle $\d_i= 2\p (1- n)$ (i.e. an opening angle which is a multiple of $2\p$).  Such singularities have been argued to correspond to the insertion of a heavy degenerate primary in the dual CFT \cite{Castro:2011iw},\cite{Perlmutter:2012ds},\cite{Raeymaekers:2014kea}.
\item{\bf Parabolic singularity:} For $\e_i = 1/4$, the monodromy belongs to a parabolic conjugacy class, i.e. $M_i$ is not diagonalizable but can be brought in a canonical Jordan form  with a nonzero element above the diagonal. The base geometry near the defect is that of the constant negative curvature metric near a puncture.
\item{\bf Hyperbolic singularity:} For $\e_i > 1/4$  the monodromy  is hyperbolic, i.e. with real non-degenerate eigenvalues.  A defect with hyperbolic monodromy creates a hole in the base manifold \cite{Seiberg:1990eb}, and the metric is the constant negative curvature metric on the cylinder.
From the dual CFT point of view, these solutions mimic a left-moving  thermal ensemble \cite{Fitzpatrick:2014vua},\cite{Fitzpatrick:2015zha} with temperature
$t_L = \sqrt{4\e_i-1}/2\p $.
\end{itemize}

\paragraph{Parameter counting}

Let us now count the number of real constraints on $c_j$ that we get if we require the monodromy matrices $M_i$ to be in $SU(1,1)$. Let us fix  the values of the $c_j$ parameters arbitrarily and let us determine what is the dimension of the space in which the monodromy matrices around the insertion points in the unit disk take values. Each monodromy matrix in this situation takes its value in $SL(2,\mathbb{C})$ which gives us $6N$ parameters. But these parameters are not completely independent: the trace of the monodromy matrices is fixed in terms of the coefficients of double poles in $T$ which reduce the number of parameters by $2N$. Furthermore, as we show in Appendix \ref{appA}, by the reflection condition on $f$ the total monodromy around the boundary of the disk is always in $SU(1,1)$ independently of $c_j$ and this reduces the number of parameters by $3$. Finally, we have an overall $SU(1,1)$ freedom of conjugating all the monodromy matrices by a constant $SU(1,1)$ matrix. Generically this conjugation has no stabilizer so we need to subtract another $3$ real parameters to describe the space of possible monodromy matrices up to conjugation. In total, we find the dimension of this space to be $6N-2N-3-3 = 2(2N-3)$ real dimensional.

In order to find the number of real conditions on $c_j$, we now compute the dimension of the space of monodromy matrices after imposing the $SU(1,1)$ conditions. An $SU(1,1)$ monodromy matrix around a given point in  the disk has $3$ real parameters, so we start with $3N$ real parameters. The trace of the monodromy matrix is now automatically real, but the double poles fix the values of this real trace and so this reduces the number of degrees of freedom by $N$. The condition that the monodromy around the boundary of the disk is in $SU(1,1)$ is now automatic, because all the individual monodromies already lie in $SU(1,1)$. Finally, we fix the overall $SU(1,1)$ conjugation freedom as before, reducing the number of parameters by another $3$. In total, the dimension of the space of allowed monodromy matrices up to global $SU(1,1)$ conjugation is $3N-N-3=2N-3$ dimensional.

In order to get from general $SL(2,\mathbb{C})$ values of monodromy matrices to $SU(1,1)$ we need to tune the accessory parameters $c_j$ by requiring $2(2N-3)-(2N-3) = 2N-3$ conditions which is exactly the number of independent real $c_j$'s. This parameter counting thus shows us that the accessory parameters are locally uniquely determined by the requirement of $SU(1,1)$ monodromy.

For comparison, we can also do a similar parameter counting in the case of the sphere. The general space of $SL(2,\mathbb{C})$ monodromy matrices up to conjugation has dimension $6n-2n-6-6=2(2n-6)$. $2n$ comes from the fact that the (complex) traces are fixed from the double poles. One of the factors of $6$ is a consequence of $M_1 M_2 \cdots M_n = 1$ condition on monodromy matrices on the sphere and the other factor of $6$ is result of quotienting out our moduli space by global $SL(2,\mathbb{C})$ transformations. The space of $SU(1,1)$ monodromy matrices on the sphere has dimension $3n-n-3-3=2n-6$ where again $N$ comes from the real trace, $3$ comes from the $M_1 M_2 \cdots M_n = 1$ condition and finally $3$ is a result of fixing the global $SU(1,1)$ freedom. Their difference is $2n-6$ which is again equal to the number of independent accessory parameters on the sphere.

\subsection{Relation to classical conformal blocks}\label{blockeq}

There is another related and well-studied monodromy problem \cite{Zam0} (see \cite{Harlow:2011ny},\cite{Hartman:2013mia},\cite{Litvinov:2013sxa} for reviews) which is that of classical conformal blocks. Quantum conformal blocks $\mathcal{F}(z_i,h_i,\Delta_j)$ are the basic holomorphic building blocks for correlation functions in conformal field theory. Restricting to the case of spherical conformal blocks, these are holomorphic (not necessarily single-valued) functions of $n \geq 4$ points $z_i$ on a sphere and depend in addition on the fusion channel, on the central charge of the theory, on conformal dimensions $h_i$ of primary operators inserted at points $z_i$ (external dimensions) and on $(n-3)$ conformal dimensions $\Delta_j$ of primary operators whose conformal families are exchanged (internal dimensions). See the figure \ref{fig2} for an example.

\paragraph{Classical conformal blocks on the sphere} The classical conformal blocks are obtained from the quantum conformal blocks by taking a scaling limit $c\to\infty$ while keeping the classical conformal dimensions $\epsilon_i \sim h_i/c$ and $\nu_i \sim \Delta_i/c$ fixed. In this limit, the rescaled logarithm of the quantum conformal block is expected to have a finite limit and this is the classical conformal block
\begin{equation}
\mathcal{F}(z_i,h_i,\Delta_j) = \exp \left[ -\frac{c}{6} {\rm f}(z_i,\epsilon_i,\nu_j) \right].
\end{equation}

It can be shown \cite{Litvinov:2013sxa}  that the classical conformal blocks can be found by solving the following monodromy problem: consider an ODE
\begin{equation}
\left(\partial_z^2 + t(z)\right)\psi(z) = 0
\label{odecb}
\end{equation}
with
\begin{equation}
t(z) = \sum_{k=1}^n \left( \frac{\epsilon_k}{(z-z_k)^2} + \frac{d_k}{z-z_k} \right).
\end{equation}
The parameters $\epsilon_k$ are the classical conformal dimensions and $d_k$ are the accessory parameters. In order to avoid an additional singularity at $z=\infty$, the accessory parameters must satisfy
\begin{eqnarray}
\sum_k d_k & = & 0 \\
\sum_k (d_k z_k + \epsilon_k) & = & 0 \\
\sum_k (d_k z_k^2 + 2\epsilon_k z_k) & = & 0.
\end{eqnarray}
For fixed values of positions $z_i$ and classical dimensions $\epsilon_k$ the number of complex independent accessory parameters $d_k$ is $n-3$. These accessory parameters are fixed by requiring the following: the monodromy matrix of the solutions of ODE around a curve that encircles the insertion points of primaries that have fused should have trace equal to
\begin{equation}
-2\cos(\pi \sqrt{1-4\nu})
\end{equation}
where $\nu$ is the classical conformal dimension of the fused primary field whose internal line is intersected by the curve in the conformal block diagram. See figure \ref{fig2} for an example
 that will be relevant for us. There are $n-3$ such conditions, one for each internal line of the conformal block diagram, so by parameter counting we can expect that this fixes the accessory parameters locally uniquely. The accessory parameters determined in this way are related to the classical conformal block ${\rm f}(z_k,\epsilon_k,\nu_j)$ through the equation
\begin{equation}
d_k = \frac{\partial}{\partial z_k} {\rm f}(z_l,\epsilon_l,\nu_j).\label{accpar}
\end{equation}
In other words, knowing the classical conformal block, we can compute from it the values of accessory parameters such that the solutions of ODE (\ref{odecb}) have prescribed conjugacy class of monodromy matrices around $n-3$ non-intersecting cycles that are specified by the fusion channel of the corresponding conformal block (see figure \ref{fig2}).

\paragraph{Connecting two monodromy problems} To connect this to our $SU(1,1)$ monodromy problem, we make a simple but important observation, proven in  Appendix \ref{appA}, which is that the reflection condition (\ref{reflf}) implies that if the monodromy around singularity $z_i$ is $M_i$, the monodromy around its mirror image point $1/\bar{z}_i$ is
\begin{equation}
M_{\bar{i}} = \sigma_3 M_i^\dagger \sigma_3\label{monodrim}
\end{equation}
(we are encircling the singular point $1/\bar{z}_i$ in the same counterclockwise direction and choosing the same basepoint $p$ on a boundary as explained in the Appendix \ref{appA}).  In particular, if $M_i \in SU(1,1)$, we have
\begin{equation}
M_i^\dagger \sigma_3 M_i = \sigma_3
\end{equation}
so in this case
\begin{equation}
M_{\bar{i}} = M_i^{-1},
\end{equation}
i.e. the monodromy around a singularity is the inverse of the one around its mirror image. From this it follows that if we compute the monodromy matrix around a simple curve that encircles pairs of singularities and their mirrors, we will get a trivial monodromy, i.e. the monodromy matrix corresponding to identity exchange, $\nu=0$. Conversely, imposing that the monodromies around curves encircling mirror pairs of singularities are trivial and assuming a $\ZZ_2$  symmetry which leads to the reflection property (\ref{reflTL}) of the stress tensor, we can ensure by an overall $SL(2, \CC)$ transformation that (\ref{monodrim}) holds, which leads to
 \be
1= M_{\bar{i}} M_i= \sigma_3 M_i^\dagger \sigma_3 M_i
 \ee
or in other words, the monodromies around each of the $z_i$ lie within $SU(1,1)$.

This implies that our $SU(1,1)$ monodromy problem on the disk with $N$ insertion points is a special case of the classical conformal block monodromy problem on a sphere with $n=2N$ punctures inserted in the points $z_i$ and their images $1/\bar z_i$, in the channel that is shown in figure \ref{fig2}. All the exchanged conformal families are those of the identity operator $\nu=0$. Assuming knowledge of the classical conformal blocks (considering ${\rm f}(z_k,\epsilon_k,\nu_j)$ as a known special function), we have determined the accessory parameters for our $SU(1,1)$ monodromy problem, finally reducing the solution of the Liouville equation to integrating the ODE (\ref{LiouvFuchs}).

\subsection{Example 1: one elliptic singularity on the disk}\label{secex1}
Let's illustrate the general construction of the solutions discussed in this section to the example of single elliptic defect in the unit disk at $z=z_0$ with $\delta = 2\pi(1-a)$. Here $a$ is the geometric opening angle around $z_0$ in units of $2\pi$. The reflection property (\ref{reflTL}) determines the accessory parameters uniquely and we find
\be
T  = {(1 - a^2) ( z_0 - {1 \over \bar z_0})^2\over 4 (z-z_0)^2  ( z - {1 \over \bar z_0})^2}.
\ee
The two linearly independent solutions can be chosen as
\begin{eqnarray}
\nonumber
\psi_1 & = & \frac{1}{\sqrt{a(1-|z_0|^2)}} (z-z_0)^{\frac{1+a}{2}} (1-\bar{z}_0 z)^{\frac{1-a}{2}} \\
\psi_2 & = & \frac{1}{\sqrt{a(1-|z_0|^2)}} (z-z_0)^{\frac{1-a}{2}} (1-\bar{z}_0 z)^{\frac{1+a}{2}}
\label{psi1center}
\end{eqnarray}
(the normalization is such that the Wronskian is $1$) and their ratio is
\begin{equation}
f = \frac{\psi_1}{\psi_2} = \left( \frac{z-z_0}{1-\bar{z}_0 z} \right)^a.\label{off-center}
\end{equation}
We already chose these two solutions such that $f(z)$ satisfies the reflection condition (\ref{reflf}). The monodromy matrix around the singular point $z = z_0$ is $\rm{diag}(e^{\pi i(1+a)}, e^{\pi i(1-a)})$.

For $z_0=0$, we expect this solution  to describe a  a conical defect  in the center of global AdS$_3$, and we will now check  this explicitly. After plugging (\ref{off-center}) into  (\ref{ourframe}) , (\ref{ansatz})and (\ref{Asol}) with $b=1$, we find that the coordinate transformation
\bea
z &=& (\tanh \r )^{1/a} e^{i (\t/a + \f)}\\
t &=& \t
\eea
brings the metric in the form
\be
ds^2 = l^2 \left[-\cosh^2 \r d\t^2 +  d\r^2 + a^2 \sinh^2 \r d\f^2\right].
\ee
Since $\f$ has period $2\p$, this is indeed the  metric of  conical defect with deficit angle $\delta = 2\pi(1-a)$ in the center of AdS$_3$.

\subsection{Example 2: two elliptic singularities, perturbative in second charge}\label{secex2}
As a second example, we consider the case  of two elliptic singularities, $\d_1, \d_2$, where $\d_2 \ll \d_1$, so that we can do perturbation theory on the
background with deficit angle $\d_1$. This problem was considered in the context of Liouville theory on the pseudosphere in \cite{Menotti:2006gc}. For simplicity we put the heavy source with $\delta_1 = 2\pi(1-a)$ at $z=0$ and the light source with $\delta_2 = 4\pi \epsilon$ at $z = r \in \RR$. The stress tensor then has the expansion
\bea
T  &=& T_0 + \e T_1 + \calo (\e^2 )\\
T_0 &=& {1-a^2 \over 4 z^2}\\
T_1 &=& {1\over (z-r)^2 }+{1\over (z-1/r)^2} + {c_0 \over z} + {c_r \over z-r} + {c_{1\over r} \over z-1/r}.\label{accesspar}
\eea
(we used the fact that all accessory parameters in our problem start at order $\epsilon$ and we rescaled them by $\epsilon$ relative to the general discussion before). Using the reflection property of the stress-energy tensor (\ref{reflTL}) we can solve for $c_0, c_{1\over r}$ and find
\be
T_1 = {\left(r - {1\over r}\right)^2 \over (z-r)^2\left( z-{1\over r}\right)^2} + {2r - c_r (1-r^2)\over z(z-r)\left( z-{1\over r}\right)}.
\ee
and furthermore $c_r \in \mathbb{R}$. We expand the solutions to the Fuchsian differential equation (\ref{Fuchs1}) as
\be
\Psi = (1+\epsilon B)(\Psi^0 + \e  \Psi^1)
\ee
where $\Psi^0$ is the unperturbed solution (\ref{psi1center}) with $z_0=0$, and $\Psi^1$ satisfies
\be
 {\Psi^1} '' + T_0 \Psi^1 = - T_1 \Psi^0.\label{eq2center}
\ee
We have also made use of the freedom to multiply $\Psi$ by a matrix $1 + \epsilon B$ in $SL(2,\CC)$, which, as we expect from our discussion in appendix \ref{appA}, will be  needed in order
to satisfy the ZZ boundary condition (\ref{ZZpsi}).

The solutions to equation (\ref{eq2center}) are of the form \cite{Menotti:2006gc}
\bea
\psi^1_i &=& M_i^{j}(z) \psi_j^0\\
M_i^j &=& \e^{jk} \int_0^z dx \psi_i^0 (x) \psi_k^0(x)  T_1(x), \qquad \e^{12}\equiv 1\label{Mmat}
\eea
These integrals can be evaluated explicitly and the result is
\bea
M_1^1&=& - M_2^2 = {1\over a} \left[ {2-z(r +r^{-1}) \over (z-r)(z-r^{-1})} + (1+rc_r) \ln {z-r \over r^2(z-r^{-1})} - 2 \right]\\
\nonumber
M_1^2 &=& {z^{1+a} \over a} \left[ {-2 z+(r +r^{-1}) \over (z-r)(z-r^{-1})} - {1+a+rc_r\over (1+a)r}\,_2F_1(1+a,1,2+a,z r^{-1})\right.\\ && \left. + {(1-a+rc_r)r\over 1+a}\,_2F_1(1+a,1,2+a,z r ) \right]\\
\nonumber
M_2^1 &=& {z^{1-a} \over a} \left[ {2 z-(r +r^{-1}) \over (z-r)(z-r^{-1})} + {1-a+rc_r\over (1-a)r}\,_2F_1(1-a,1,2-a,z r^{-1})\right.\\ && \left. - {(1+a+rc_r)r\over 1-a}\,_2F_1(1-a,1,2-a,z r ) \right].
\label{Mcomps}\eea
The final solution for $\Psi$ is therefore to first order in $\epsilon$
\be
\Psi = (1+ \epsilon B) (1 + \e M) \Psi^0.
\ee
The requirement which will fix $c_r$ is that the monodromy around $z=r$ belongs to the subgroup $ SU(1,1) \subset SL(2,\CC)$. The change in $\Psi$ as we encircle $z=r$ comes purely from the change in the matrix elements $M_i^j$ (since $\Psi^0$ was regular at $z = r$):
\be
\d \Psi =\e \d M  \Psi + \calo( \e^2).
\ee
Therefore the matrix $B$ does not contribute to the monodromy at this order; it will instead be fixed by the ZZ boundary condition (\ref{Liouvbc}). From (\ref{Mmat}) we see that the jump $\d M_i^j $  comes from a contour integral around $z=r$
\be
\d M_i^j =  \e^{jk} \oint_r dx \psi_i^0 (x) \psi_k^0(x) \d T_1(x).
\ee
Computing the residue of the integrand at $z=r$ leads
to
\be
\d M_i^j = 2\p i \e^{jk} \left( c_r \psi^0_i (r) \psi^0_k (r) + \psi^{0\prime}_i(r)  \psi^0_k(r) + \psi^0_i(r) \psi^{0\prime}_k(r) \right).
\ee
Explicitly one obtains
\bea
\d M_1^1 &=& - \d M_2^2 = {2\p i \over a} (1 + r c_r)\\
\d M_1^2 &=&  {-2\p i r^a \over a} (a+1+ r c_r) \\
\d M_2^1 &=&  {-2\p i r^{-a} \over a} (a-1- r c_r).
\eea
Imposing that the monodromy is in $ SU(1,1) \subset SL(2,\CC)$ means that we should require
\bea
\tr \d M &=& 0\\
\d M_2^2 &=& \overline{\d M_1^1}\\
\d M_2^1 &=& \overline{\d M_1^2}.
\eea
The first condition is automatically satisfied, the second one is automatically satisfied for real $c_r$ (which already followed from the reflection property of $T $), and the last one determines $c_r$ to be
\be c_r =-{1\over r} \left( 1+a {r^a + r^{-a}\over r^a - r^{-a}}\right).\label{crsol}
\ee

It remains to verify that it is possible to adjust the constant $SL(2,\mathbb{C})$ matrix $B$ such that $\Psi(z)$ satisfies the boundary condition (\ref{ZZpsi}). For this  we need to find a matrix $B$ such that the reflection property (\ref{reflPsi}) is satisfied. In terms of matrices $M$ and $B$ this reduces to condition
\begin{equation}
M(z) - \sigma_1 \bar{M}(1/z) \sigma_1 = -B + \sigma_1 \bar{B} \sigma_1
\end{equation}
or
\begin{eqnarray}
M_1^1(z) - \bar{M}_2^2(1/z) & = & -B_1^1 + \bar{B}_2^2 \\
M_1^2(z) - \bar{M}_2^1(1/z) & = & -B_1^2 + \bar{B}_2^1.
\end{eqnarray}
In particular, the left-hand side of these expressions should be $z$-independent and the right-hand side has $SU(1,1)$ freedom undetermined as we expect from the general discussion. As soon as we show that the LHS is $z$-independent, we can always find matrix $B$ which satisfies these two equations, even without having an explicit expression for the LHS. The $z$-independence follows directly from the integral representation of $M_i^j(z)$ and reflection property of one-defect solutions,
\begin{eqnarray}
\frac{d}{dz} \left(\bar{M}_2^2(1/z) \right) & = & \frac{1}{z^2} \bar{\psi}_1^0(1/z) \bar{\psi}_2^0(1/z) \bar{T}_1(1/z) = \psi_1^0(z) \psi_2^0(z) T_1(z) = \frac{d}{dz} M_1^1(z) \\
\frac{d}{d z} \left( \bar{M}^1_2(1/z) \right) & = & -\frac{1}{z^2} \bar{\psi}_2^0(1/z) \bar{\psi}_2^0(1/z) \bar{T}_1(1/z) = -\psi_1^0(z) \psi_1^0(z) T_1(z) = \frac{d}{d z} M_1^2(z).
\end{eqnarray}
For completeness, we can evaluate the integrals explicitly with result
\begin{eqnarray}
M_1^1(z) - \bar{M}_2^2(1/z) & = & -\frac{2}{a}\left(1 + (1+r c_r)\log r\right) \\
M_1^2(z) - \bar{M}_2^1(1/z) & = & \frac{(-1)^a \pi}{a \sin(\pi a)}\left[(1+rc_r)(r^a-r^{-a}) + a(r^a+r^{-a}) \right].
\end{eqnarray}
The second expression vanishes when the accessory parameter $c_r$ takes the correct value, so we see that the matrix $B$ can be chosen to be diagonal, with simplest choice being
\begin{equation}
B = \frac{1}{a}\left[1+(1+r c_r)\log r\right] \sigma_3.
\end{equation}

Before ending this section, let us verify our main result of section \ref{blockeq} for this example, namely that the accessory parameter is determined by a classical vacuum conformal block in the channel illustrated in figure \ref{fig2}. This conformal block was computed, in the same perturbative approximation as in our current example, in \cite{Hijano:2015rla}.
To compare with that paper, it's useful to make a scale transformation $\tilde z = r z$, such that the singularities are in $0,1,\infty$, and define  $x\equiv r^2$. From the conformal transformation of the stress tensor and the expansion (\ref{accesspar}) it's easy to see  that under such a  rescaling, the accessory parameter transforms as $\tilde c_x = {c_r \over r}$, leading to
 \be
 \tilde c_x = - {1\over x} \left( 1+a {x^{a\over 2} + x^{-{a\over 2}}\over x^{{a\over 2}} - x^{-{a\over 2}}}\right).
 \ee
Comparing with (2.225) in \cite{Hijano:2015rla} this is precisely value of the accessory parameter obtained when computing the vacuum block.

\section{The holographic stress tensor}\label{sechol}
In the previous section we used concepts from conformal field theory as a tool  to argue for the existence of certain gravity solutions, without making direct reference to the AdS/CFT correspondence. In this section we will explore a little more what can be said about the interpretation of our solutions in  a holographically dual CFT.
We will compute
the holographic stress tensor for our solutions describing point particles moving on helical Lorentzian geodesics in the bulk,   using the standard holographic dictionary. This will confirm the  picture we arrived at with heuristic arguments in section \ref{secansatz}.

We start by deriving the near-boundary behaviour of the metric (\ref{ansatz}). For this, we will need
 to know the first nonvanishing subleading term in the near-boundary expansion (\ref{Liouvbc}) of the Liouville field.
As we show in Appendix \ref{appas}, there is no term of order $(1-|z|^2)^3$  in $e^{2\F}$   and we have
\be
e^{2 \F} = { (1-|z|^2)^2} + f_2 (\arg z) (1-|z|^2)^4 + \calo( (1-|z|^2)^5 )\label{Liouvbc2}
\ee
where $f_2 (\arg z)$ is a completely arbitrary function, undetermined by the Liouville equation. We also show in Appendix \ref{appas} that the  subleading terms in (\ref{Liouvbc2})
are uniquely determined in terms of $f_2$ through recursion relations.

The free function $f_2 (\arg z)$ contains the same information as the Liouville stress tensor $T $ defined in  (\ref{TLiouv}):
 indeed, substituting the expansion (\ref{Liouvbc2}) in (\ref{TLiouv}), we find that $f_{2}$ is essentially the boundary value of $T $:
 \be
 f_2(\arg z) =\left. -\left( {z^2\over 3} T (z)\right)\right\vert_{|z|=1}.\label{f2TL}
 \ee
 It's important to note that, since $f_2$ is real, the right hand of (\ref{f2TL}) side must be real. This proves, as promised, the  reality condition on the stress tensor (\ref{realTL}) which we assumed so far.

We will now use these observations to derive an expression for the holographic stress tensor \cite{Henningson:1998gx},\cite{Balasubramanian:1999re}.
For this, we need to bring the metric in a form  which manifestly obeys the Brown-Henneaux falloff conditions, which in Fefferman-Graham coordinates look like
\be
ds^2 = l^2\left[ {dy^2\over 4 y^2} - {dx_+ dx_- \over y} + {6\over c} T^{cyl}(e^{i x_+}) dx_+^2 + {6\over c} \tilde T^{cyl}(e^{i x_-}) dx_-^2+\calo(y  )\right]\label{asadsmetr}
\ee
where $c= 3l/2 G$, $y$ is a radial coordinate such that the boundary is at $y=0$,  and  $ x_\pm = \t \pm \f $ where $\f$ is an angular variable with period $2 \p$.
The arbitrary functions $T^{cyl}(e^{i x_+}), \tilde T^{cyl}(e^{i x_-})$ are the VEVs of the $++$ and $--$ components of the stress tensor in the dual CFT,  defined on the cylinder.
The zero-modes of $T^{cyl}$ and $\tilde T^{cyl}$ are related to mass $M$ and angular momentum $J$ and to the left- and rightmoving conformal weights $h,\bar h$ as
\bea
T^{cyl}_0 = \half (Ml + J)= -{c\over 24 } + h \\ \tilde T^{cyl}_0=\half(Ml-J)= -{c\over 24 } + \bar h. \label{Tconst}
\eea
For example, global AdS corresponds to $M = -1/8G, J=0$  or $T^{cyl} = \tilde T^{cyl} = - c/24$, and represents the left-and right moving ground state of the dual CFT.

We will now derive the boundary stress tensor for the  solutions (\ref{ansatz}) obeying the boundary condition (\ref{Liouvbc}).
We substitute the expansion (\ref{Liouvbc2}) and the expression for $A$ in (\ref{Asol}) into  (\ref{ansatz}) and make the coordinate transformation
\bea
z &=& \left( 1- { 1- 4 b G m_{tot}\over 2} y + { (1- 4 b G m_{tot})^2 \over 8} y^2 \right)e^{i x_+} + \calo (y^3 )\\
t &=& \half \left(x_+ + (1- 4 b G m_{tot}) x_- - 2 b i \l (e^{i x_+} )\right) + \calo(y^2).\label{toBH}
\eea
Here, $m_{tot}= \sum_{i=1}^N m_i$ and we recall that $b$ was a constant introduced in (\ref{Asol}) and which should be taken to be 1 in the current context (the case $b=0$ will be discussed below).
One can check using (\ref{lambdadef}) that the coordinates $x_\pm$ indeed have the periods $(x_+, x_-)\sim (x_++ 2\p, x_--2\p)$.
The metric is now of the form (\ref{asadsmetr}) with
\be
T^{cyl}(e^{i x_+}) = - {c\over 24}+ {c \over 6} e^{2 i x_+} T  (e^{ i x_+})  ,\qquad \tilde T^{cyl} (e^{i x_-}) = -{c\over 24} + \bar h  \label{Tcyl}
\ee
where the right-moving weight is
\be
\bar h =   {b \over 2} m_{tot} l (1-2 b G m_{tot}).\label{hbar}
\ee
We note that the right-moving stress tensor only has the zero mode turned on, while left-moving stress tensor is determined by the Liouville stress tensor in the bulk.
Setting $b=1$ and using (\ref{accesspar1}) and (\ref{Tcyl}) we find for the mass and angular momentum of our solutions
\bea
M l &=& -{c\over 12} + 2 \bar h + J\\
J &=& {4 c G^2 \over 3}\sum_{i<j} m_i m_j +{c \over 6} \sum_i c_i z_i .\label{MJsols}
\eea
It is important to remark that  the coordinate transformation (\ref{toBH}) is valid only when $m_{tot}$ lies below the upper bound
\be
m_{tot} \leq {1 \over 4 G} .
\ee
Since in this range we have $Ml \leq J$, the meaning of this bound is that our solutions obey the Brown-Henneaux falloff conditions only when the total mass and angular momentum are outside of  BTZ black hole regime.

Let us discuss the result (\ref{MJsols}) in some examples. As a first check, we note that for a single particle of mass $m$ in $z=0$, we obtain
\be
M  =-{l\over 8 G} + m(1-2 mG), \qquad J=0. 
\ee
This reproduces the standard result for the mass of the backreacted point mass solution in the center of  AdS$_3$ (see e.g. \cite{David:1999zb}), with the term linear in $G$ representing gravitational interaction energy. For a single particle of mass $m$ in $z=z_0$, as in the example in \ref{secex1}, we find instead
\bea
M &=& -{1\over 8 G} + {m(1-2 mG)\over 1- |z_0|^2 } \\
J &=& {m l(1-2 mG)|z_0|^2\over 1- |z_0|^2 }. 
\eea
For $|z_0|>0$, this solution approaches an extremal spinning BTZ black hole for $m \to {1 \over 4 G}$. For the two-center example of \ref{secex2}, we find, in the notation introduced there,
\bea
Ml &=&  -{c a^2\over 12} + {a \e c\over 3} +J + \calo (\e^2)\\
J&=& -{c\over 6}a \e\left( 1+ {r^a + r^{-a}\over r^a - r^{-a}}\right) .
\eea

As we saw in (\ref{Tcyl}), the left-moving boundary stress tensor  $T^{cyl}$ is closely related to the holomorphic Liouville stress tensor $T(z)$.  We can gain more insight into this relation
by considering the stress tensor VEV for the Wick-rotated Euclidean CFT defined on the plane, whose holomorphic and anti-holomorphic parts we will denote by  $T^{pl}(u)$ and $ \bar T^{pl}(\bar u)$ respectively.
To obtain them, we first analytically continuing $\t \to - i \t_E$, which sends $i x_+ \to w, ix_- \to \bar w$ with $w$ a complex coordinate on the cylinder and subsequently apply the conformal map $u = e^{ w}$
 from the cylinder to the plane  parametrized by $u$.
 We find, using the standard conformal transformation of the CFT stress tensor,
\be
T^{pl}(u) ={T^{cyl}(u)+ {c\over 24} \over u^2}, \qquad \bar T^{pl}(\bar u) ={\tilde T^{cyl}(\bar u)+ {c\over 24} \over \bar u^2}. \label{Tplane}
\ee
For our solutions, using (\ref{Tcyl}), we obtain simply
\be
T^{pl}(u) = {c \over 6} T (u), \qquad \bar T^{pl}(\bar u ) ={\bar h \over \bar u^2}  \label{TitoTl}.
\ee
with $\bar h$ given in (\ref{hbar}).
Interestingly, the Liouville stress tensor $T  (z)$, which is a bulk quantity constructed from the metric, coincides, when analytically continued from the unit disk to the plane,  with
the holomorphic stress tensor of the dual CFT.
 As we  saw in the previous section, the analytically continued Liouville stress tensor $T  (u)$ has double pole singularities
at the   locations $z_i$ of the particles and at their image points $1/\bar z_i$. 
Therefore,  in the dual CFT on the plane, purely holomorphic  operators are inserted at the $z_i$ and  their image points $1/\bar z_i$. From the expression for the anitholomorphic stress tensor we see that there are purely anti-holomorphic operator insertions in the origin  $u=0$ and at $u = \infty$. These observations confirm the picture we had arrived heuristically in section \ref{secansatz}.

We now discuss the interpretation of taking the parameter $b$, introduced in  (\ref{Asol}), to zero. In that case, the one-form $A$ has Dirac string singularities
and the Einstein equations are now solved with different delta-function sources. From (\ref{hbar}),(\ref{TitoTl}), we see that in this limit $T^{pl} (u)$ is unchanged while
$\bar T^{pl} (\bar u) \to 0$. Hence these solutions describe purely holomorphic operator insertions in the dual CFT, and are in some sense extremal since they
saturate the unitarity bound $\bar h \geq 0$. The corresponding short representations of the bosonic $SL(2, \RR) \times \overline{ SL(2, \RR) }$ symmetry algebra of AdS$_3$ are  called singletons \cite{Flato:1990eu}. One expects them to have similar  properties to BPS states in supersymmetric theories, and  indeed one can show that the metric (\ref{ansatz}) possesses a timelike Killing spinor which is antiperiodic on the boundary cylinder\footnote{One should distinguish such extremal particle solutions, which saturate the unitarity  bound
$|J| \leq Ml + {c \over 12}$   from extremal black hole solutions, which instead saturate the bound $|J|\leq M l $
 expressing the existence of a horizon, and allow for a Killing spinor which is periodic (i.e. in the R sector) \cite{Coussaert:1993jp}.}
 (i.e. in the NS sector of the dual CFT) \cite{Izquierdo:1994jz}.

It is therefore plausible to conjecture that the $b=0$ solutions  arise from  sources describing extremal spinning particles
with $m_i = j_i$. Since spinning particle sources are most easily described in the Chern-Simons description of 3D gravity \cite{Witten:1989sx}, we can make this more precise by
working out the Chern-Simons gauge fields for our solutions.
    From the vielbein and spin connection we can construct  two  flat $SL(2,\RR) $  connections $\cala,\tilde \cala$ as follows:
\be
\cala = \left(\o^a + {1 \over l} e^a\right) J_a, \qquad
\tilde \cala = \left(\o^a - {1 \over l} e^a\right) J_a.
\ee
where\footnote{Our conventions for the $sl(2,\RR)$ Lie algebra are $[J_a,J_b] = \e_{ab}^{\ \ c} J_c$ with
$\e_{012} = -1$ and indices are lowered with $\h_{ab} = {\rm diag} (-1,1,1)$. For definiteness we will use the two-dimensional representation
$J_0 = {i \over 2} \s_3, J_1=\half \s_1, J_2 = \half \s_2$ for which $\tr J_a J_b = \half \h_{ab}$.} $\o^a \equiv \half \e^a_{\ bc} \o^{bc}$. Choosing the vielbein
\be
e^0 = {l } ( dt + A), \qquad
e^1 =  {l } e^{- \F} \Re e dz,\qquad
e^2 =  {l } e^{- \F} \Im m dz
\ee
and computing the corresponding spin connection, we obtain
\bea
 \cala &=& 2 \Im m (\pa_z \F d z) J_0  +  e^{-\F} dz J_-  +  e^{-\F} d\bar z J_+\label{Laxconn}\\
\tilde \cala &=& -2 ( dt + b d (\Im \l) ) J_0
\eea
where $J_\pm = J_1 \pm i J_2$ and $\l$ was defined in (\ref{lambdadef}). This shows that $ \cala$ is time-independent and is the standard
Lax connection whose flatness implies the Liouville equation, see e.g. \cite{Babelon}, ch. 12. Singularities in the Liouville field   show up in the Chern-Simons formulation as monodromy defects of $\cala$ (see also \cite{Harlow:2011ny}), i.e. singularities around which $\cala$ has nontrivial monodromy.
While for nonzero $b$ the connection $\tilde \cala$ has singularities due to the multivaluedness of $\l$,
for $b \to 0$ the connection $\tilde \cala$ becomes pure gauge. Therefore this situation corresponds to coupling extremal spinning particles which  source only the leftmoving field  $\cala$. Note also that $\cala$ and $\tilde \cala$ are presented in a rather different gauge than the standard Fefferman-Graham type gauge of \cite{Banados:1998ta}.

\section{On solutions with abelian monodromy}\label{secabel}
In our construction of multi-centered solutions so far, it was of great importance to allow the monodromy group associated with the Fuchsian differential  equation  (\ref{Fuchs1}) to be a {\em nonabelian} subgroup of $SU(1,1)$. In this section, we will discuss whether it is possible to construct solutions where the monodromy group is actually abelian. As we shall see, this is actually not possible without
introducing extra spurious singularities in the differential equation, which as discussed in section \ref{secmon} correspond to extra pointlike  sources with negative masses.  As reviewed there, these sources do have a dual CFT interpretation as describing insertions of degenerate primaries \cite{Castro:2011iw},\cite{Perlmutter:2012ds}\cite{Raeymaekers:2014kea}. The corresponding solutions  can be simply written down analytically.
   This observation also leads to a new understanding of the problems which arise when one attempts to construct  static solutions as we will discuss in section (\ref{secstatic}) below.

\subsection{Unremoveable spurious  singularities}
We would like to answer the question if we can construct solutions where the monodromy matrix around each of the sources $z_i$ is a diagonal matrix of the form diag $\left(e^{{i\d_i\over 2}},e^{{-i\d_i\over 2}}\right)$.
It follows from our discussion of parameter counting in the monodromy problem that this imposes $2(N-1)$ more real conditions than there are accessory parameters in $T $.
  It is known in the literature on Fuchsian differential equations (see e.g. \cite{Yoshida}) that these extra conditions can be met by   introducing $ (N-1)$ spurious singularities in the unit disk, whose positions provide the sought-after extra parameters. The symmetry of the problem dictates that we will have an additional $N-1$  spurious singularities in the corresponding image points.

  The resulting solutions can be written down analytically and result from multiplying together the functions $f$ for
single-center solutions in (\ref{off-center}):
\be
f = \prod_{i=1}^N \left( {z - z_i \over 1- \bar z_i z}\right)^{a_i}\label{fnaive}
\ee
where $a_i = 1- { \d_i \over 2 \p}$.
One easily checks that these satisfy the boundary conditions and solve the field equations with elliptic singularities in the points $z_i$. However, they also contain additional
spurious singularities. From the discussion of section \ref{secmon}, these arise from zeroes of $f'$ which are not zeroes of $f$, where the metric has a conical excess of a multiple of $2\p$. Computing $f'/f$ from (\ref{fnaive}) we find
\be
{f' \over f}(z) = {P(z) \over Q(z)}
\ee
with
\be
P(z) = \sum_{i=1}^N a_i (1- |z_i|^2) \prod_{j\neq i}(z-z_i)(1-\bar z_i z),\qquad
Q(z)= \prod_{i=1}^N (z-z_i)(1 - \bar z_i z).
\ee
Hence the spurious singularities are located in the zeroes of $P(z)$, which is generically a polynomial of order $2(N-1)$ in $z$.

A natural question to ask is whether we can remove the spurious singularities by tuning the  parameters $a_i, z_i$ in such a way that $P(z)$ becomes a nonvanishing constant. This turns out not to be possible, except for the trivial  single-centered $N=1$ case.
Indeed, one easily sees that the coefficients in the expansion
\be
P(z) = \sum_{n=0}^{2 (N-1)} p_n z^n
\ee
satisfy
\be
p_n = \bar p_{2(N-1)-n}.
\ee
Therefore, if we tune the parameters such that the coefficient of $z^{2(N-1)} $ vanishes, automatically also the constant term $p_0$ vanishes and therefore $P(z)$ cannot be made a (nonzero) constant.

It's easy to show that if $s$ is a zero of $P(z)$, then so is its image $1/\bar s$, so $P(z)$ can be written as
\be
P(z) = p_{2 (N-1)} \prod_{j =1}^{N-1} (z- s_j) \left(z-{1 \over \bar s_j}\right)
\ee
where $p_{2 (N-1)}$ and $s_i$ are functions of $a_i, z_i$ and the $s_i$ are located in the unit disk. The Liouville field on the disk constructed from $f$  satisfies
\be
\pa_z \pa_{\bar z}\F + e^{-2\F } = \half \sum_{i = 1}^N \d_i \d^{(2)} ( z- z_i,\bar z - \bar z_i)- \p \sum_{j = 1}^{N-1}  \d^{(2)} ( z- s_j,\bar z - \bar s_j).
\ee
In the dual CFT, this will describe $N$ pairs of nondegenerate primaries with weights $h_i = {c \over 24}(1-a_i^2)$ inserted in $z_i$ and $1/\bar z_i$ respectively, and typically $N-1$ pairs of degenerate primaries $\calo_{(1,2)}$
with weight $-c/8 +\calo(1)$ inserted at the $s_j$ and  $1/\bar s_j$.
If some of the $s_j$ are multiple zeroes, we will get less than $N-1$ pairs of  degenerate primaries, and a zero of  order $m$ leads to an insertion of
 a pair of  primaries  $\calo_{(1,m+1)}$ with weight $-{c\over 24}(1-(m+1)^2) +\calo(1)$.

\subsection{Static multicenter solutions revisited}\label{secstatic}
Here we will revisit static solutions with multiple particle sources,
originally studied in \cite{Deser:1983nh}. We will show that such solutions are also described by
solutions of the Liouville equation with sources, but that there are additional constrains which force the monodromy group to be abelian.
From the analysis above, we know that such solutions must contain additional spurious singularities. In this way we recover, in our framework, the
observations in the literature \cite{Clement:1994qb},\cite{Coussaert:1994if},\cite{Mansson:2000sj} concerning the presence of extra singularities in static multi-centered solutions.

We start from our general ansatz (\ref{genKV}), setting $A=0$ to obtain a static metric. Setting $\L = -1/l^2$, rescaling the coordinates by the AdS radius $l$ and shifting  $\F$ by  $- \ln 2$ for convenience,
the metric looks like
\be
ds^2= l^2 \left[ -N^2 dt^2  + 4 e^{-2\F} dz d\bar z  \right]
\ee
where $N(z,\bar z), \F(z, \bar z)$. The Einstein equations (\ref{Einst1}-\ref{Einst4}) become
\bea
\pa_z \left( \pa_z N e^{2\F}\right) &=&0\label{statEinst1}\\
 \pa_z\pa_{\bar z} N - 2 N e^{-2\F}   &=&0\label{statEinst2}\\
 \pa_z\pa_{\bar z}\F + e^{-2\F}  &=& 16 \p G \sum_i m_i \d^2(z- z_i, \bar z- \bar z_i).\label{statEinst3}
\eea
As before, we can argue that $z$ can be taken to run over the unit disk, and that the Liouville field $\F$ should satisfy the ZZ boundary condition (\ref{Liouvbc}) on the boundary circle.
The solution to the Liouville equation (\ref{statEinst3}) can once again be written in the form
\be
e^\F = \Psi^\dagger \s_3 \Psi
\ee
where $\Psi$ is determined up to an $SU(1,1)$ transformation.
Substituting this into (\ref{statEinst1}-\ref{statEinst2}) one finds that the  general solution for $N$ can be expressed in terms of $\Psi$ as follows
\be
N = {\Psi^\dagger C \Psi \over \Psi^\dagger \s_3 \Psi}
\ee
where $C$ is a Hermitian $2 \times 2$ matrix of integration constants:
\be
C = \left( \begin{array}{cc} r & c\\ \bar c & r\end{array} \right)\qquad r\in \RR, c\in \CC.
\ee
One can show that, by transforming $\Psi$ by an $SU(1,1)$ transformation, we can set $c$ to zero, and by a rescaling of the time coordinate we can set $r$ to one,
so that $C$ is the identity matrix. The metric then takes the form
\be
ds^2 =  {l^2 \over (\Psi^\dagger \s_3 \Psi)^2} \left[ - (\Psi^\dagger  \Psi)^2 dt^2  + 4  dz d\bar z  \right].
\ee
The observation we want to make is that the full metric is not invariant under $SU(1,1)$, but only under a $U(1)$ subgroup (generated by $i \s_3$).
In particular, the monodromies around the singular points must all lie within this $U(1)$ and therefore commute. As we showed in the previous subsection, this inevitably leads to the introduction of
additional spurious singularities. This explains  the observations made in the literature \cite{Clement:1994qb},\cite{Coussaert:1994if},\cite{Mansson:2000sj} on the presence of additional singularities in static multi-center solutions from our point of view.
\section{Outlook}
In this work we argued for the existence of multi-centered solutions describing point masses on helical worldlines  in Minkowskian AdS$_3$ using a connection to conformal blocks in Euclidean 2D CFT. We end by listing some open issues and possible generalizations:
\begin{itemize}
\item Given the fact that our bulk gravity solutions are completely determined by a Liouville field, it would be interesting to understand if the bulk action (\ref{einstact}), supplemented by suitable regularizing boundary terms,
  can be related to the regularized Liouville action  in the presence of point-like sources. For a single particle in the center of AdS$_3$, such a relation was explored in \cite{Krasnov:2000ia}.
\item In this work, we focused exclusively on  solutions with point-particle sources, around which the monodromy of $f(z)$ is elliptic. It would be interesting to generalize our results
to black-hole-like extremal solutions  where one allows also  singularities  of parabolic or hyperbolic type\footnote{Multi-black hole solutions with several asymptotic regions were studied in the literature, including \cite{Brill:1995jv}-\cite{Sheikh-Jabbari:2016unm}.}.
\item It would be interesting to generalize our results to 3D higher spin gravity, relating multi-centered  higher spin solutions to $W$-algebra conformal blocks \cite{deBoer:2014sna}.     Since massless higher spin fields are  most simply formulated in terms of Chern-Simons fields, this would involve a generalization of the
    Chern-Simons description of our solutions discussed at the end of section \ref{sechol}.
This  is currently under investigation \cite{inprog}.
\end{itemize}

\section*{Acknowledgements}
We would like to  thank  S. Konopka, E. Martinec, D. Van den Bleeken and B. Vercnocke for useful comments and discussions, and H. Maxfield for pointing out an error in v1.  The research of OH and JR was supported by the Grant Agency of the Czech Republic under the grant 14-31689S. The research of TP was supported by the DFG Transregional Collaborative Research Centre TRR~33 and the DFG cluster of excellence Origin and Structure of the Universe.

\begin{appendix}

\section{$SL(2,\CC)$ transformations and  the ZZ boundary condition}\label{appA}
In this Appendix, we will prove the following theorem:\\
 {\em Given a function $T(z)$ satisfying the reflection property (\ref{reflTL}):
 \be
 T (z) = z^{-4} \bar T  (1/z)\label{reflTapp}
 \ee
 there exists a solution $f(z)$ to the Schwarzian differential equation
 \be
 S(f,z)= 2 T(z)\label{schweqapp}
 \ee
 which satisfies the reflection condition (\ref{reflf}):
 \be
 f(z) = {1 \over \bar f(1/z)}.\label{reflfapp}
 \ee  }
We will prove this theorem by starting from an arbitrary solution to (\ref{schweqapp}) and applying a suitable $SL(2,
\CC )$ symmetry of the equation  (\ref{schweqapp}) to obtain a solution which satisfies (\ref{reflfapp}).

 First, we note that (\ref{reflTapp}) is a necessary property which follows from applying the Schwarzian derivative to both sides of (\ref{reflfapp}) and using (\ref{schweqapp}).
 Now, suppose $g(z)$ is a solution to $S(g,z)= 2T (z)$. Then it follows from (\ref{reflTapp}) that so is the function $1/\bar g(1/z)$.
 We now recall that any two solutions $f_1,f_2$ to (\ref{schweqapp}) must be related by an $SL(2,\CC)$ fractional linear tranformation\footnote{The proof follows from using (\ref{psiitof}) to show that $f_1$ and $f_2$ define
 two bases of independent solutions to the linear differential equation $(\pa_z^2 + T (z) ) \psi =0$ with  Wronskian equal to one. These two bases must be related  by an $SL(2,\CC)$ transformation, from which the stated relation between $f_1$ and $f_2$
 follows.}.
Therefore  $g(z)$ and  $1/\bar g(1/z)$ must be related by a fractional linear transformation:
 \be
 {1 \over \bar g(1/z)} = G \cdot g(z)\label{frlinrel}
 \ee
 where $G$ is and $SL(2,\CC )$ matrix, and the fractional linear action is defined as
 \be G \cdot g(z)\equiv {G_{11} g(z) + G_{12}\over G_{21} g(z) + G_{22}}.\ee
 Note that we can rewrite  (\ref{frlinrel}) as
 \be
 \bar g(1/z) = I \cdot G \cdot g(z) \label{frlinrel2}
 \ee
 where $I$ is the inversion
 \be
 I = \left( \begin{array}{cc} 0& 1\\1 &0 \end{array}\right).
 \ee
 Taking the complex conjugate of (\ref{frlinrel2}) and substituting back in (\ref{frlinrel2}) we see that $G$ is not a generic element of $SL(2,\CC )$ but satisfies
 \be
 I \bar G I G=1=  I G I \bar G.
 \ee
 For such elements one can show that they can be written as
 \be
 G = I \bar \L^{-1} I \L
 \ee
 with $\L$ another $SL(2,\CC )$ element. Substituting in (\ref{frlinrel}) we get
 \be
 {1 \over \bar \L \cdot \bar g(1/z)} = \L \cdot g(z).
 \ee
 Setting
 \be
 f(z) = \L \cdot g(z)
 \ee
 we have obtained a solution to (\ref{schweqapp}) which satisfies the reflection condition (\ref{reflfapp}).

The reflection property (\ref{reflfapp}) implies an important property of the monodromy of $f$ which is used in the main text.
Indeed, it is straightforward to show that (\ref{reflfapp}) implies that if $C$ is a closed curve and $M_C$ the monodromy of $f$ when encircling it, then the monodromy around the image $\tilde C$ of $C$ \footnote{Here by image curve $\tilde C$ we mean pointwise image of $C$, so in particular if $C$ is a simple curve in the complex plane with counterclockwise orientation, $\tilde C$ will have clockwise orientation.} under the map $z \to 1/\bar z$ is:
\be
M_{\tilde C} = \s_3 (M_C^\dagger)^{-1} \s_3.\label{monrel}
\ee
We are interested in monodromy of $f$ along a contour that encircles counterclockwise both the singular point $z_i$ and its mirror image $\bar{z}_i^{-1}$ (and no other singularities). To compute this monodromy, we pick a basepoint $p$ on a boundary of the unit disk and consider a contour $C_i$ which goes along a straight line from $p$ to a small neighbourhood of $z_i$, encircles $z_i$ counterclockwise and comes back to $p$ along the same line. We denote the monodromy transformation along this contour by $M_i$. The mirror contour $\tilde{C}_i$ has an opposite orientation, so using (\ref{monrel}) we see that the monodromy transformation around $\tilde{C}_i^{-1}$ (which still has basepoint $p$ and is oriented counterclockwise) is
\begin{equation}
M_{\bar{i}} = \sigma_3 M_i^{\dagger} \sigma_3
\end{equation}
and so the total monodromy around a symmetric contour based at at $p$ and encircling both $z_i$ and $\bar{z}_i^{-1}$ counterclockwise is
\begin{equation}
M_{\bar{i}} \cdot M_i = \sigma_3 M_{i}^{\dagger} \sigma_3 M_{i}.
\end{equation}
For $M_{i} \in SU(1,1)$ this is equal to identity matrix.

Another consequence used in the main text comes from applying  (\ref{monrel}) to a curve  $C$ which is  a circle approaching the boundary of the unit disk from the inside (still based at point $p$ on a boundary). If we have $N$ singularities inside the disk, $M_C = \prod_{i=1}^N M_{i}$. Then $\tilde C$  is a curve approaching the boundary from the outside, but traversed in the opposite direction. Applying (\ref{monrel}) we find
\be
M_C =\s_3  M_C^{\dagger} \s_3.
\ee
Therefore, if $N-1$ monodromies, say $M_{1}, \ldots, M_{N-1}$, are in $SU(1,1)$, then so is $M_{N}$.

\section{Near-boundary expansion of the Liouville field}\label{appas}
In this Appendix we consider the near-boundary expansion of a Liouville field  obeying the ZZ boundary condition (\ref{Liouvbc}). We will show that the Liouville equation can be solved recursively in a near-boundary perturbation series. Setting
\be
z = \sqrt{1- u} e^{i\f}
\ee
let's assume an expansion around $u=0$ of the form
\be
e^{2 \F} = \sum_{n=0}^\infty f_n (\f) u^{n+2}.\label{nbexp}
\ee
We now substitute this ansatz in the Liouville equation (\ref{Liouv}) without source terms. We are assuming that there are no sources on the boundary itself, so we can only hope our solution to be valid
for values of small enough values of $u$ outside of the sources. We obtain the following system of equations
\bea
0 &=& 2 (f_k - f_{k-1}) + \sum_{m=0}^k \left( (k-2m-1)(k-m+2)  f_mf_{k-m} - (2k-4m-3)(k-m+1) f_m f_{k-m-1}\right.\nonu
& &\left. +(k-m)(k-2m-2) f_m f_{k-m-2} + {1\over 4} ( f_m f_{k-m-2}'' -f_m' f_{k-m-1}')\right).
\eea
For $k=0$, this leads to $f_0=0$ or $f_0=1$, and the ZZ boundary condition instructs us to choose the latter. The $k=1$ equation then demands the first subleading term to vanish,
\be
f_1 = 0.
\ee
The $k=2$ equation is then automatically obeyed, in particular it does not put any restrictions on the function $f_2 (\f)$.
The remaining equations then express the $f_{k>2}$ recursively in terms of the arbitrary function $f_2$, through the  relations
\bea
f_k &=& {1 \over (k+1)(k-2)} \left( (1-k(5-2k))f_{k-1} - (k-2)^2 f_{k-2} - {1\over 4} f_{k-2}''\right.\nonu
&&\left.  - \sum_{m=2}^{k-2}  (k-2m-1)(k-m+2)  f_mf_{k-m} +\sum_{m=2}^{k-3} (2k-4m-3)(k-m+1) f_m f_{k-m-1}\right.\nonu
& &\left. -\sum_{m=2}^{k-4}\left(  (k-m)(k-2m-2) f_m f_{k-m-2} + {1\over 4} ( f_m f_{k-m-2}'' -f_m' f_{k-m-1}')\right) \right).
\eea

\end{appendix}

\end{document}